\documentclass[12pt,preprint]{aastex}

\slugcomment{Not to appear in Nonlearned J., 45.}

\shorttitle{AGN's FEEDBACK AND SMBH ACCRETION}
\shortauthors{Wang, J}

\begin{document}


\title{A DIRECT LINKAGE BETWEEN AGN OUTFLOWS IN THE NARROW-LINE REGIONS AND THE X-RAY EMISSION FROM THE ACCRETION DISKS}


\author{J. Wang\altaffilmark{1,2}, D. W. Xu\altaffilmark{1,2} and J. Y. Wei\altaffilmark{1,2}}
\email{wj@bao.ac.cn}
\altaffiltext{1}{Key Laboratory of Space Astronomy and Technology, National Astronomical Observatories, Chinese Academy of Sciences}
\altaffiltext{2}{National Astronomical Observatories, Chinese Academy of Sciences}







\begin{abstract}

The origin of outflow in narrow-line region (NLR) of active galactic nucleus (AGN) is studied in this paper by 
focusing on the relationship between the [\ion{O}{3}]$\lambda$5007 line profile
and the hard X-ray (in a bandpass of 2-10 keV) emission from the central SMBH in type-I AGNs. A sample of 47 local 
X-ray selected type-I AGNs at $z<0.2$ is extracted from the 2XMMi/SDSS DR7 catalog that is originally
crossmatched by Pineau et al. The X-ray luminosities in an energy band from 2 to 10keV  
of these luminous AGNs range from 
$10^{42}$ to $10^{44}\ \mathrm{erg\ s^{-1}}$.  A joint spectral analysis is performed on their optical and 
X-ray spectra, in which the [\ion{O}{3}] line profile is modeled by a sum of several Gaussian functions
to quantify its deviation from a pure Gaussian function. 
The statistics allows us to 
identify a moderate correlation with a significance level of 2.78$\sigma$: luminous AGNs with       
stronger [\ion{O}{3}] blue asymmetry tend to have steeper hard X-ray spectra. 
By identifying a role of $L/L_{\mathrm{Edd}}$ on the correlation at a $2-3\sigma$ significance level 
in both direct and indirect ways, we argue that 
the photon index versus asymmetry correlation provides
evidence that the AGN's outflow commonly observed in its NLR
is related with the accretion process occurring around the central SMBH, which favors the wind/radiation model 
for the origin of the outflow in luminous AGNs.

\end{abstract}


\keywords{galaxies: nuclei - quasars: emission lines - X-ray: galaxies}



\section{INTRODUCTION}

It is now generally believed that the feedback from central active galactic nucleus (AGN) plays  
an important role in the galaxy evolution issue 
(see reviews in Fabian 2012; Veilleux et al. 2005 and Heckman \& Best 2014). A self-regulated super-massive 
black hole (SMBH) growth and star formation in the host galaxy
is potentially realized by sweeping out circumnuclear gas
through the feedback process in both galaxies merger and secular evolution scenarios (e.g.,
Alexander \& Hickox 2012; Kormendy \& Ho 2013).
Both semianalytic models and numerical simulations indicate that 
this feedback process is useful not only for reproducing the observed
$M_{\mathrm{BH}}-\sigma_*$ relation, luminosity functions of quasars
and normal galaxies (e.g., Fabian 1999; Di Matteo et al. 2005, 2007; Hopkins et al. 2007, 2008;
Menci et al. 2008; Silk \& Rees 1998; Haehnelt et al. 1998; Khalatyan et al. 2008;
Granato et al. 2004; Somerville et al. 2008; Kauffmann \& Haehnelt 2000; Springel et al. 2005; Croton et al. 2006),
but also for solving the ``over cooling'' problem in the $\Lambda$ cold dark matter ($\Lambda$CDM) galaxy
formation model in which the cooling predicted in galaxy groups and clusters are stronger than the 
observed one  (e.g., Ciotti \& Ostriker 2007; 
Somerville et al. 2008; Hirschmann et al. 2013). 
By using the SDSS spectroscopic survey, Wang et al. (2011) and Wang (2015) recently suggest a
co-evolution between the feedback and host galaxy based on a revealed fact that an AGN with stronger blue 
asymmetry of the [\ion{O}{3}]$\lambda$5007 emission line tends to be associated with a younger stellar population.


So far, various methods of the feedback process have been proposed in past decades by a mixture 
of observational and theoretical studies. These methods include  
AGN's wind (e.g., Crenshaw et al. 2003; Pounds
et al., 2003; Ganguly et al. 2007; Reeves et al. 2009; Dunn et al.,
2010; Tombesi et al., 2012), radiation pressure (e.g., Granato et al. 2004; Alexander
et al. 2010) and mechanical energy outflow caused by collimated radio jet (e.g., Best et al. 2006; Rosario et al. 2010; Holt et al.
2008; Nesvadba et al. 2008; Guillard et al. 2012).

On the observational ground, there is accumulating evidence that the feedback from central SMBH can drive outflows
on various scales (see reviews in Veilleux et al. 2005 and Fabian 2012). 
In addition to the blueshifted absorption lines in optical, UV and soft X-ray spectra (e.g., Crenshaw et al. 2003; Hamann \& Sabra 2004;
Wang \& Xu, 2015 and references therein), the outflows can be conveniently traced by the blue asymmetry 
of the [\ion{O}{3}]$\lambda\lambda$4959, 5007 doublet and its bulk blueshift
with respect to the local system (Heckman et al. 1981; Veron-Cetty et al. 2001; Zamanov et al. 2002;
Xu \& Komossa 2009; Marziani et al. 2003; Aoki et al. 2005; Boroson 2005; Bian et al. 2005; Komossa et al. 2008; 
Zhang et al. 2013; Mullaney et al. 2013).

Using the [\ion{O}{3}] line profile as a diagnostic of the outflow occurring in AGN's narrow-line region (NLR) enables
the studies of the origin of the outflow (feedback) based on large optical spectroscopic sample.
Different results are, however, obtained by various authors. On the one hand, there is ample evidence that
the [\ion{O}{3}] line blue asymmetry and bulk velocity blueshift are strongly correlated with Eddington ratio 
($L/L_{\mathrm{Edd}}$, where $L_{\mathrm{Edd}}=1.26\times10^{38} (M_{_\mathrm{BH}}/M_\odot)\ \mathrm{erg\ s^{-1}}$ is
the Eddington luminosity) of the central AGNs: higher the $L/L_{\mathrm{Edd}}$, stronger the blue asymmetry and larger 
the bulk velocity blueshift will be (e.g., Zhang et al. 2011; Wang et al. 2011; Wang 2015; Boroson 2005; Bian et al. 2005).
In addition, the extreme ``blue outliers'' usually defined as the objects with [\ion{O}{3}] bulk blueshift larger
than $250\mathrm{km\ s^{-1}}$ (e.g., Zamanov et al. 2002; Zhou et al. 2006; Komossa et al. 2008) are found to 
exclusively occur in the AGNs associated with high $L/L_{\mathrm{Edd}}$. 
On the other hand, the relation 
between the [\ion{O}{3}] line profile and radio emission has been reported in the previous studies. 
Early studies show that the [\ion{O}{3}] line width is found to be correlated with radio
luminosity at 1.4 GHz ($L_{\mathrm{1.4GHz}}$) for a sample of flat-spectrum radio galaxies (e.g., Heckman et al. 1984; Whittle 1985).
Mullaney et al. (2013) and Zakamska \& Greene (2014) recently analyzed the [\ion{O}{3}] line profile in a large sample of both type I and type II AGNs 
detected by the SDSS spectroscopic survey. Their results suggest that the [\ion{O}{3}] line width 
is more strongly related to $L_{\mathrm{1.4GHz}}$ than the other AGN’s parameters (e.g., line luminosity and $L/L_{\mathrm{Edd}}$).

The origin of feedback in local type I AGNs is studied in this paper by focusing on the relationship between 
the outflow traced by the [\ion{O}{3}] line profile and the X-ray emission from the central AGNs. 
AGNs are well known to be luminous X-ray emitters up to 100 keV.
X-ray emission is a powerful tool for identifying nuclear SMBH accretion activity
and for studying the accretion process that fuels AGNs, because the luminous X-ray emission from a rapidly accreting 
AGN is produced
in the region very close to the SMBH (e.g., Haardt \& Maraschi 1991; Zdziarski et al. 2000; Kawaguchi et al.
2001; Cao 2009).

The paper is organized as follows. The sample selection and spectral analysis are presented in \S 2 and 
\S 3, respectively. The statistical results are shown in \S 4, and   
the implications are discussed in \S 5. A $\Lambda$CDM cosmology
with parameters $H\mathrm{_0=70\ km\ s^{-1}\ Mpc^{-1}}$, $\Omega_{\mathrm m}=0.3$, and $\Omega_{\Lambda} = 0.7$ 
is adopted throughout the paper.

\section{SAMPLE SELECTION: Type-I AGNs from the 2XMMi/SDSS-DR7 Catalog}

The sample of X-ray selected type-I AGNs used in the current study is extracted from the 2XMMi/SDSS-DR7 catalog, which was
originally crossmatched by 
Pineau et al. (2011) between the incremental Second XMM-Newton Serendipitous Source
Catalog (2XMMi, Watson et al. 2009) and the SDSS-DR7 catalog (Abazajian et al. 2009). 
In the crossmatch, the optical counterpart of an X-ray source is identified by  
the probability of spatial coincidence estimated from the traditionally adopted likelihood 
ratio estimator. There are in total 221,012 unique, serendipitous X-ray sources in
the 2XMMi catalog. A $\sim$90\% completeness can be achieved for the catalog at 
a sensitivity of $1\times10^{-14}\ \mathrm{erg\ s^{-1}\ cm^{-2}}$ and $9\times10^{-14}\ \mathrm{erg\ s^{-1}\ cm^{-2}}$ 
in the 0.5-2.0 keV and 2.0-12.0 keV bandpasses, respectively. The localization accuracy of 
the X-ray sources is typical of $2\arcsec$.
More than 30,000 X-ray point-like sources with a 
localization accuracy $\geq5\arcsec$ have a SDSS-DR7 optical counterpart
with an identification probability larger than 90\%.

A sub-sample of X-ray type-I AGNs is selected from the 2XMMi/SDSS-DR7 catalog 
by requiring that: 
(1) the probability of an identification is no less than 95\%, (2) the angular distance between
an individual XMM-Newton X-ray source and its corresponding
optical counterpart is less than 3\arcsec, taking into account 
the SDSS fiber aperture, (3) the redshift is smaller than 0.2, (4) the g-band brightness is 
brighter than 19 mag, which is necessary for a proper modeling of the optical continuum, and (5) X-ray flux in the 0.2-12 keV bandpass is larger than
$1\times10^{-14}\ \mathrm{erg\ s^{-1}\ cm^{-2}}$, which excludes the extracted X-ray spectra with low
photon count rates. By focusing on the objects that are classified as quasars according
to the spectral type classification given by the SDSS pipelines
(Glazebrook et al. 1998; Bromley et al. 1998), there are a total of 82 type-I AGNs fulfilling
the above selection criteria, after removing SDSS\,J093249.57+472522.8 
which shows a spectrum typical of a CV star and is incorrectly classified as a quasar by
the SDSS pipelines (Yip et al. 2004).  By examining the SDSS spectra one by one by eyes,  19 out of the 82 sources are then excluded 
from the subsequent optical and X-ray spectral analysis, 
because of their Seyfert-1.8 like spectra (i.e., partially obscured AGNs) 
in which the continuum is dominated by the starlight from their host galaxies.  

\section{DATA REDUCTIONS}

\subsection{XMM-Newton EPIC Spectra}

Our X-ray spectral analysis focus on the XMM-Newton EPIC PN (Struder et al. 2001) data.
The data are reduced by the SAS v11.0 software\footnote{http://xmm.esac.esa.int/} and by the corresponding calibration
files. The events with patterns of 0-4 are extracted from the PN data for all the 63 X-ray type-I AGNs selected in 
Section 2, except for one object.
Single pixel events (i.e., pattern=0) are extracted for one bright object (SDSS\,J103438.59+393828.2=KUG\,1031+398) since 
these events are less sensitive to pile-up than other patterns. 
In the extraction, the bad and hot pixels are
removed from the original image, and the CCD chip gaps avoided.
The source spectrum
is extracted from a circular aperture at the detected source
position. The aperture has a radius of 25\arcsec-40\arcsec\ depending on
the brightness of the object. The background is determined from
a circular source-free region that is offset from, but close to,
the source. The pile-up in the data is checked by the SAS task
epatplot. The tasks rmfgen and arfgen are used to generate the
needed response files.

There are in total 47 X-ray selected type-I AGNs for
the subsequent X-ray spectral modelings. The other 16 objects are excluded either because of the 
coincidence of the CCD gap or because of the bad X-ray spectral quality beyond 2-3 keV due to their faintness.  
We fit the extracted spectra over the 0.3-10 keV band 
by the XSPEC package (Arnaud 1996). 
The absorption caused by our own galaxy is included in 
the spectral fitting for each object by taking the column density value from  
the Leiden/Argentine/Bonn (LAB) Survey (Kalberla et al.
2005). A basic model expressed as 
\it wabs*zwabs*(N$\times$blackbody+powerlaw)\rm\ is adopted in the fitting for all of the 
objects. The power-law photon spectrum is defined as $N(E)\propto E^{-\Gamma}$, where $E$ is the
photon energy and $\Gamma$ is the photon index.
We also attempt to reproduce each of the spectra by the neutral reflection model (\it pexrav\rm; Magdziarz \& Zdziarski 1995)
instead of the simple powerlaw, although no significant improvement (i.e., the difference of the reduced $\chi^2$ is less than 10\%)
can be obtained from this complicate model.
In addition to the basic model, additional components are required for some objects to reproduce the 
observed spectra. A Gaussian profile is required in three objects (SDSS\,J091848.61+211717.0, SDSS\,J140700.40+282714.6 = Mark\,668, and  
SDSS\,J143452.45+483942.7 =NGC\,5683) to model their broad iron K$\alpha$ emission lines at 6.4 keV (rest frame).
At low energy end, the 
X-ray spectra of SDSS\,J091848.61+211717.0, SDSS\,J111706.39+441333.3 (PG\,1114+445) and SDSS\,J124210.61 +331702.6 (WAS\,61)
are best fitted by an edge-like absorption (\it zedge\rm) due to the K-edge of ions. 
The best-fit edge energy $E_{\mathrm{edge}}$ in rest frame is $0.67\pm0.04$, $0.72\pm0.01\ \mathrm{keV}$ and 
$0.73\pm0.01\ \mathrm{keV}$ for SDSS\,J091848.61+211717.0, SDSS\,J111706.39+441333.3 and SDSS\,J124210.61+331702.6, respectively.
These modeled edge energies 
suggest that the  edge-like absorptions in the three objects are likely caused by the K-shell absorption of \ion{O}{7} ions at 0.74keV.
Our best fit model of PG\,1114+445
is highly consistent with that was obtained from the \it ROSAT\rm\ PSPC observation (Laor et al. 1994).  
As an illustration, Figure 1 shows the X-ray spectral modelings for four objects\footnote{We have checked the resulted photon indices 
by modeling the X-ray spectra in the 2-10keV bandpass with a simple model of \it wabs*zwabs*(powerlaw+Gaussian). \rm The sample model returns 
consistent photon indices within their uncertainties when compare with the values obtained from the fittings based on the 
0.3-10 KeV band.}.

\subsection{SDSS Optical Spectroscopy}

We analyze the one-dimensional optical spectra of the 47 X-ray selected type-I AGNs by the IRAF\footnote{IRAF is distributed by 
National Optical Astronomy Observatory, which is operated by the Association of Universities for Research in Astronomy, Inc.,
under cooperative agreement with the National Science Foundation.} package as follows. 
At the beginning, the correction of Galactic extinction is applied to each spectrum 
according to the color excess, the parameter $E(B-V)$ taken
from the Schlegel, Finkbeiner, and Davies Galactic reddening
map (Schlegel et al. 1998). An extinction law of the MilkyWay with $R_{\mathrm V}=3.1$ (Cardelli et al. 1989) is
adopted in the correction. Each of the spectra is then
transformed to the rest frame, along with the flux correction due
to the relativity effect given the redshift provided by the SDSS
pipelines.

\subsubsection{Continuum modeling and removal}

The continuum of each rest-frame spectrum is modeled by the linear sum of 
several components, they are 1) a broken powerlaw from central AGN, 
in which the wavelength of the break point and the two spectral indices are 
not fixed in the continuum modeling, 2) a template of both high order Balmer emission lines and a Balmer continuum
from the broad-line region (BLR), 
3) an empirical template of the optical \ion{Fe}{2} complex, and 4) the eigenspectra of the host galaxy that are built 
from the standard single stellar population spectral library through the 
principal component analysis (PCA) method (e.g., Hao et al. 2005; 
Wang \& Wei 2008; Francis et al. 1992). 
The used spectral library is developed by Bruzual \& Charlot (2003).
A Galactic extinction curve with $R_V=3.1$ is involved in the modeling to account for the intrinsic 
extinction due to the host galaxy. 
A $\chi^2$ minimization is iteratively performed
over the rest-frame wavelength range from 3700\AA\ to 7500\AA, except
for the regions with strong emission lines. Figure 2 illustrates the modeling and
removal of the continuum for two typical objects.

The template of the high order Balmer lines (i.e., $\mathrm{H_7-H_{50}}$)
is taken from the 
case B recombination model with electron temperature of $T_e=1.5\times10^4\mathrm{K}$ and electron 
density of $n_e=10^{8-10}\ \mathrm{cm^{-3}}$. The model is calculated by
Storey \& Hummer (1995). The line width of the high order Balmer lines is fixed to be that of the 
broad component of H$\beta$, which is determined by our line profile modeling (see below).    
The Balmer continuum is approximately modeled by the emission from a partially  
optically thick cloud by following  Dietrich et al. (2002, see also in Grandi 1982; Malkan \& Sargent 1982):
\begin{equation}
 f_\lambda^{\mathrm{BC}}=f_\lambda^{\mathrm{BE}}B_\lambda(T_e)(1-e^{-\tau_\lambda}); \lambda\leq\lambda_{\mathrm{BE}}
\end{equation}
where $f_\lambda^{\mathrm{BE}}$ is the continuum flux at the Balmer edge $\lambda_{\mathrm{BE}}=3646$\AA, $B_\lambda(T_e)$ is 
the Planck function at an electron temperature\footnote{Malkan \& Sargent (1982) shows that 
the Balmer continuum of AGNs is best fitted with a temperature of $T_e\sim1.5\times10^4$K in the optically thin case or 
$T_e\sim10^3$K in the optically thick case. } of $T_e=1.0\times10^4$K, and $\tau_\lambda$ is the optical depth at wavelength $\lambda$.
The optical depth at $\lambda$ is related with the one at the Balmer edge $\tau_{\mathrm{BE}}$ as $\tau_\lambda=\tau_{\mathrm{BE}}(\lambda/\lambda_{\mathrm{BE}})^3$.
$\tau_{\mathrm{BE}}$ ranges from 0.1 to 2.0, and 
a typical value of $\tau_{\mathrm{BE}}=0.5$ is adopted in the current Balmer continuum fitting.

We model the optical \ion{Fe}{2} complex by using the empirical templates provided in Veron-Cetty et al. (2004). 
Both the broad and narrow components of the \ion{Fe}{2} templates are included in the modeling.   
Again, the line widths of the broad and narrow \ion{Fe}{2} emission are determined from the 
line profile modeling of the H$\beta$ emission line (see below). With the modeled \ion{Fe}{2} complex, its
flux (i.e., FeII$\lambda$4570) is measured from the modeled spectrum
in the rest-frame wavelength range from 4434 to 4684\AA.
 
\subsubsection{Emission line profile modeling}

After the removal of the continuum, the emission-line profiles are modeled on each emission-line isolated 
spectrum for both H$\alpha$ and H$\beta$ regions (i.e., in the wavelength ranges $\lambda\lambda$6350-6750 and 
$\lambda\lambda$4800-5050) by the SPECFIT task (Kriss 1994)
in the IRAF package. For each object, each emission line is profiled by a linear combination of 
a set of several Gaussian functions. The line flux ratios of the [\ion{O}{3}] and [\ion{N}{2}] doublets are 
fixed to their theoretical values. 
The line width of the narrow H$\alpha$ (H$\beta$) component is fixed to equal to that of [\ion{N}{2}] ([\ion{O}{3}]
core) line, if the resulted two widths are different significantly.   
In order to properly isolate their [\ion{O}{3}]$\lambda5007$ line profile (see Section 3.1 below),
the broad \ion{He}{1}$\lambda5016$ emission line (Veron et al. 2002) is additionally
required in the line profile modelings in three objects 
(i.e., SDSS\,J111830.28+402554.0, SDSS\,J134022.86+274058.5 and SDSS\,J155909.62+350147.4).

The line modelings are
schematically presented in the left and right panels in Figure 3 for the H$\alpha$ and H$\beta$ regions, respectively.
As shown in the figure, a linear combination of two or three broad Gaussian functions is usually required to 
adequately reproduce the observed broad Balmer line profile in most cases. 
A residual line profile, which is obtained by subtracting the modeled narrow line component (including the modeled
forbidden lines) from the observed profile, is then used to measure the line width and integrated 
line flux of either broad H$\alpha$ or H$\beta$ emission line. 
Figure 4 schematically shows the [\ion{O}{3}]$\lambda$5007 line profiles modeled by a sum of $n$ Gaussian functions
for four typical cases. As illustrated by the figure, generally speaking, a sum of two or three Gaussian functions are adequate to 
model all the observed [\ion{O}{3}] line profiles well.





\section{ANALYSIS AND RESULTS}

Table 1 lists the results obtained from the above X-ray and optical spectral modelings for the 47 
X-ray selected type-I AGNs. 
The identification of each object and the corresponding
redshift given by the SDSS pipelines are listed in Columns (1) and (2), respectively. 
The full width at half maximum (FWHM) of the broad H$\beta$ emission is tabulated 
in Column (3). Columns (4) and (12) lists the modeled broad H$\alpha$ emission line luminosity ($L_{\mathrm{H\alpha}}$) and the 
spectral-fitting-inferred intrinsic 
X-ray luminosity in a bandpass of 2-10keV ($L_{2-10\mathrm{keV}}$), respectively. 
The calculated line luminosities are corrected for the local extinction that is inferred 
from the narrow-line ratio H$\alpha$/H$\beta$ by assuming a Balmer decrement for standard case B recombination and a Galactic extinction
curve with $R_V=3.1$. 
Figure 5 shows a tight relationship between $L_{\mathrm{H\alpha}}$ and 
$L_{2-10\mathrm{keV}}$, which is not related to AGN's NLR. 
A distribution of the inferred $L_{2-10\mathrm{keV}}$ is plotted in the inert panel of the figure, which shows 
a range from $10^{42}$ to $10^{44}\ \mathrm{erg\ s^{-1}}$ for $L_{2-10\mathrm{keV}}$. 
The parameter of RFe defined as the flux ratio between the \ion{Fe}{2}$\lambda$4570 and H$\beta$ broad component
is tabulated in Column (5). The fitted spectral photon index $\Gamma_{2-10\mathrm{keV}}$ 
and the corresponding errors  at a confidence level of 90\% are listed in Column (6). These errors reported by XSPEC package are
obtained from our spectral modelings.
The measured $\Gamma_{\mathrm{2-10keV}}$ has an average value of 1.86 and a standard deviation of 0.31,
which are highly consistent with the typical value of $\Gamma\sim1.9$ for 
radio-quiet AGNs (e.g., Zdziarski et al. 1995; Reeves \& Turner 2000; Piconcelli et al.
2005; Dadina 2008; Panessa et al. 2008; Zhou \& Zhang 2010; Corral et al. 2011; Mateos et al. 2010).

\begin{description}
\item \bf [\ion{O}{3}]$\lambda5007$ bulk velocity shift \rm 
Column (7) tabulates the calculated [\ion{O}{3}]$\lambda5007$ line bulk relative velocity shift defined as
$\Delta\upsilon=c\Delta\lambda/\lambda_{0,[\mathrm{OIII}]}$, where $\lambda_{0,[\mathrm{OIII}]}$ and $\Delta\lambda$ denote the
rest-frame wavelength in vacuum of the [\ion{O}{3}]$\lambda$5007 emission line and the wavelength
shift with respect to the narrow H$\beta$ line, respectively.  $\Delta\lambda$ is calculated from the modeled line centers as
$\Delta\lambda=(\lambda_{\mathrm{[OIII]}}^{\mathrm{ob}}-\lambda_{\mathrm{H\beta}}^{\mathrm{ob}})-(\lambda_{0,\mathrm{[OIII]}}-\lambda_{0,\mathrm{H\beta}})$
where $\lambda_{\mathrm{[OIII]}}^{\mathrm{ob}}$ ($\lambda_{\mathrm{H\beta}}^{\mathrm{ob}}$) and $\lambda_{0,\mathrm{[OIII]}}$ 
($\lambda_{0,\mathrm{H\beta}}$) are the observed line center and the line wavelength in vacuum of the [\ion{O}{3}] (H$\beta$) line, respectively. 
The narrow H$\beta$ line shows
a very small velocity shift relative to the galaxy rest frame (e.g., Komossa et al. 2008), 
although this point is argued against by recent studies\footnote{These results mean that, strictly speaking, 
the obtained bulk velocity shift of [\ion{O}{3}] assesses the bulk relative velocity of high ionized gas with 
respect to that of low ionized gas.} (e.g., Hu et al. 2008; Bae \& Woo 2014; Wang \& Xu 2015). 
A negative value of $\Delta\upsilon$ corresponds to a blueshift, and a
positive value to a redshift.

\item \bf [\ion{O}{3}]$\lambda5007$ line asymmetry \rm Various ``asymmetry'' indices are commonly used in previous studies to quantify 
the asymmetry of [\ion{O}{3}] emission line 
(e.g., Heckman et al. 1981; Whittle 1985; Veilleux 1991; Wang et al. 2011; Liu et al. 2013; Harrison et al. 2014). Briefly speaking,
most of these indices (e.g., $\mathrm{AI_{20}}$ and AI) give a quantified asymmetry by comparing the measured line widths/centers 
(in unit of either wavelength or velocity) at different line flux level.  
By modeling the [\ion{O}{3}]$\lambda$5007 emission line profile by a sum of several Gaussian profiles, we parametrize the asymmetry  of 
the [\ion{O}{3}] lines by a velocity off $\delta\upsilon$ defined as  
\begin{equation}
  \delta\upsilon=\frac{\sum_{k=1}^na_k(\upsilon_k-\upsilon_p)}{\sum_{k=1}^na_k}
\end{equation}
where $a_k$ and $\upsilon_k$ is the modeled flux and velocity of the $kth$ Gaussian function, respectively. 
$\upsilon_p$ denotes the velocity of the Gaussian profile that 
reproduces the peak of the observed line profile. A negative value of $\delta\upsilon$ denotes a blue asymmetry, and 
a positive one a red asymmetry.
The fitted parameters $\delta\upsilon$ are listed in Columns (10) in Table 1.
Figure 6 compares the value of $\delta\upsilon$ used in this study with the parameter of velocity offset $\delta\upsilon_{\mathrm{H14}}$ 
defined in Harrison et al. (2014). We calculate the values of $\delta\upsilon$ for all the 47 X-ray selected type I AGNs by following the 
definition in Harrison et al. (2014): $\delta\upsilon_{\mathrm{H14}}=(\upsilon_{05}+\upsilon_{95})/2$, 
where $\upsilon_{05}$ and $\upsilon_{95}$ are the velocities at the
5\% and 95\% percentiles of the overall emission-line, respectively. Although a small systematical difference,
one can see from the figure a well correlation between the two parameters.

\item \bf Uncertainty estimation \rm  Except for $\Gamma_{2-10\mathrm{keV}}$, all the quoted errors in Table 1 correspond to a 1$\sigma$ significance level.
A proper error propagation is considered in the derivation of the uncertainties for some parameters. Specifically 
speaking, the uncertainty of RFe is estimated as 
\begin{equation}
  \Delta\mathrm{RFe}=\mathrm{RFe}\sqrt{\bigg(\frac{\Delta f_{\mathrm{FeII}}}{f_{\mathrm{FeII}}}\bigg)^2+
  \bigg(\frac{\Delta f_{\mathrm{H\beta}}}{f_{\mathrm{H\beta}}}\bigg)^2}
\end{equation}
where $f_{\mathrm{FeII(H\beta)}}$ and $\Delta f_{\mathrm{FeII(H\beta)}}$ is the measured flux of the optical \ion{Fe}{2} complex 
(broad H$\beta$ emission) and the corresponding uncertainty, respectively. We estimate the final uncertainty of $\Delta\upsilon$ as
\begin{equation}
 \delta\Delta\upsilon=c\sqrt{\bigg(\frac{\Delta\lambda_{\mathrm{[OIII]}}}{\lambda^0_{\mathrm{[OIII]}}}\bigg)^2+
 \bigg(\frac{\Delta\lambda_{\mathrm{H\beta}}}{\lambda^0_{\mathrm{H\beta}}}\bigg)^2}
\end{equation}
where $\Delta\lambda_{\mathrm{[OIII](H\beta)}}$ and $\lambda^0_{\mathrm{[OIII](H\beta)}}$  is the uncertainty of the measured line center and 
the rest frame wavelength in vacuum of the [\ion{O}{3}] (narrow H$\beta$) emission line, respectively, and $c$ is the light speed.    
The uncertainty of $\delta\upsilon$ is determined through the formula
\begin{equation}
 \mathrm{std. \delta\upsilon}=\sqrt{\sum_{k=1}^{n}(\Delta^2 a_k'+\Delta^2\upsilon'_k)} (k\not = p)
\end{equation}
where $\Delta a_k'$ is the uncertainty of the weight ($=a_k/\sum a_k$) of the $kth$ Gaussian function and 
$\Delta\upsilon'_k=\sqrt{\Delta^2\upsilon_k+\Delta^2\upsilon_p}$ ($k\not = p$). $\Delta\upsilon_k$ and 
$\Delta\upsilon_k$ is the uncertainty of the velocity of the $kth$ Gaussian function and that of the 
Gaussian function that reproduces the line peak. 
\end{description}

\subsection{$\Gamma_{\mathrm{2-10keV}}$ versus [OIII]$\lambda$5007 Line Profile}

The aim of this paper is to study the relationship between the properties of AGN's X-ray emission and the strength of the 
feedback traced by the outflow in AGN's NLR. The main results are presented in Figure 7 by 
using the [\ion{O}{3}]$\lambda5007$ emission line profile as a diagnose for the outflow occurring in the NLR.
The related statistical results are listed in Table 2.

The measured hard X-ray photon index $\Gamma_{\mathrm{2-10keV}}$ is plotted against the resulted [\ion{O}{3}] line profile asymmetry parameter 
$\delta\upsilon$ in the left panel of Figure 7. One can see there is a moderate anti-correlation between the two variables. The relationship means 
that stronger the blue asymmetry of the [\ion{O}{3}]$\lambda5007$ emission line, steeper the
X-ray spectrum will be. A Spearman rank-order test 
returns a correlation coefficient of $r_s=-0.411$. The corresponding probability of null correlation from two-tailed  is calculated to be
$p_s=0.0054$, which corresponds to a significant level at $2.78\sigma$.

The right panel of Figure 7 shows a similar relationship between the X-ray photon index $\Gamma_{\mathrm{2-10keV}}$
and bulk velocity shift $\Delta\upsilon$ of the [\ion{O}{3}]$\lambda5007$ line: larger the bulk blueshift of the 
[\ion{O}{3}] line, steeper the X-ray spectrum will be. 
The two points marked with horizontal arrows at the left side of the plot are the  
two objects (SDSS\,J092247.02+512038.0 and SDSS\,J140621.89+222346.5) with 
remarkable [\ion{O}{3}] line bulk blueshifts, i.e., $\Delta\upsilon\leq -300\ \mathrm{km\ s^{-1}}$. These values allow 
us to classify the two objects as ``blue outliers'' in which the [\ion{O}{3}] bulk blueshift is defined to be larger 
than $250\mathrm{km\ s^{-1}}$ (e.g., Zamanov et al. 2002; Komossa et al. 2008). The measured $\Gamma_{\mathrm{2-10keV}}$
of the two objects are as large as $2.15\pm0.09$ and $2.23\pm0.10$.
The Spearman rank-order test yields a correlation coefficient of $r_s=-0.483$ with a probability of null correlation of
$p_s=0.0022$ (i.e., a significance level at $3.06\sigma$). The statistics is slightly degraded to $r_s=-0.426$ with $p_s=0.0087$
(i.e., a significance level at $2.62\sigma$)  when 
the two objects with extremely large bulk blue velocity shifts are excluded.

\subsection{Relation with Eigenvector-I Space}

We examine the relation between the identified $\Gamma_{\mathrm{2-10keV}}$ versus [OIII] line profile correlations and the 
well documented AGN's Eigenvector-I (EI) space in this section. 
The EI space is one of the key properties of AGN phenomena. It was first introduced by Boroson \& Green (1992, hereafter BG92) 
for a sample of 87 bright Palomar-Green quasars. In addition to the original anti-correlation between the intensities of 
the optical \ion{Fe}{2} blends and [\ion{O}{3}] emission, the
space has been subsequently extended to infrared, UV and soft X-ray bands (e.g., Wang et al. 1996;
Sulentic et al. 2000a, 2002, 2004; Xu et al. 2003, 2012; Grupe 2004;
Laor et al. 1997; Lawrence et al. 1997; Grupe et al. 1999; Vaughan
et al. 2001; Zamanov et al. 2002; Marziani et al. 2003; Marziani \& Sulentic 2012; Wang et al. 2006). 
Up to the date, the best EI space that is widely accepted involves three parameters: 
RFe, FWHM of the H$\beta$ broad component and photon index
in soft X-ray, which means the relation with EI space can be studied equivalently by 
focusing on the two parameters: RFe and FWHM(H$\beta$) in the current sample.
The measured RFe in this paper
is in fact found to be correlated with FWHM(H$\beta$). With the Spearman rank-order test, the correlation coefficient 
and corresponding probability of null correlation are 
calculated to be $r_s=-0.397$ and $\rho_s=0.0071$ (2.69$\sigma$), respectively.    

%

Figure 8 illustrates not only the RFe versus  FWHM(H$\beta$) correlation through the 
symbol size, but also the dependence of the two identified correlations on both two parameters RFe and FWHM(H$\beta$). 
The corresponding correlation coefficient matrix is listed in Table 2.    
Each correlation coefficient and the corresponding probability of the null correlations shown in bracket
are calculated through the Spearman rank-order test.
One can learn from the table that there is a significant relationship between $\Gamma_{\mathrm{2-10keV}}$  and the EI space, which 
implies that the EI space can be well reproduced in the current X-ray selected type I AGN sample, 
although the EI space is found to be marginally (moderately)  correlated with the line asymmetry index $\delta\upsilon$  
(bulk velocity shift $\Delta\upsilon$). 
A significant anti-correlation between hard X-ray photon index $\Gamma$ and FWHM of broad 
H$\beta$ emission line has been firmly established in previous studies (e.g.,  Brandt et al. 1997; Leighly 1999;
Reeves \& Turner 2000; Shemmer et al. 2006, 2008; Zhou \& Zhang 2010; Jin et al. 2012).

\subsection{Role of SMBH Mass and Eddington Ratio}

Both SMBH mass ($M_{\mathrm{BH}}$) and $L/L_{\mathrm{Edd}}$ are critical parameters describing AGN's 
phenomena (e.g., Shen \& Ho 2014). In fact, BG92 first argued that the 
EI space is potentially driven by $L/L_{\mathrm{Edd}}$, which is 
then confirmed by various authors (e.g., Boroson 2002; Sulentic et al. 2000; Xu et al. 2003, 2012;
Marziani et al. 2003b) since the great progress made in the reverberation mapping technique
(e.g., Kaspi et al. 2000, 2005; Peterson \& Bentz 2006; and see Marziani \& Sulentic 2012 and Peterson 2013 for recent reviews). 
To investigate the role played by the two 
basic parameters in AGN's feedback process and to explore the physical origin of 
the identified two new correlations, we here at first estimate the $M_{\mathrm{BH}}$ and 
$L/L_{\mathrm{Edd}}$ for the used hard X-ray selected type-I AGNs from their optical spectra, and 
then study the statistical properties of the estimated $M_{\mathrm{BH}}$ and 
$L/L_{\mathrm{Edd}}$.

\subsubsection{Derivation of $M_{\mathrm{BH}}$ and $L/L_{\mathrm{Edd}}$} 

The calculated $M_{\mathrm{BH}}$ and $L/L_{\mathrm{Edd}}$ are shown in Columns (8) and (9) in Table 2 for each 
of the X-ray selected type-I AGNs, respectively.    
We estimate $M_{\mathrm{BH}}$ from the modeled broad H$\alpha$ line emission for all the 
47 X-ray selected type-I AGNs, except for SDSS\,J015950.24+002340.8, 
according to the calibrated relationship  provided in Greene \& Ho (2007, and references therein)
\begin{equation}
 M_{\mathrm{BH}}=(3.0^{+0.6}_{-0.5})\times10^6\bigg(\frac{L_{\mathrm{H\alpha}}}{10^{42}\ \mathrm{ergs\ s^{-1}}}\bigg)^{0.45\pm0.03}
\bigg[\frac{\mathrm{FWHM(H\alpha)}}{10^3\ \mathrm{km\ s^{-1}}}\bigg]^{2.06\pm0.06}M_\odot
\end{equation}
where $L_{\mathrm{H\alpha}}$ is the intrinsic luminosity of the H$\alpha$ broad component  
corrected for local extinction  and $\mathrm{FWHM(H\alpha)}$ is the line width of broad H$\alpha$ emission that 
is resulted from our line profile modeling (Section 3.2.2). With the estimate $M_{\mathrm{BH}}$ (i.e., the Eddington luminosity),
the Eddington Ratio $L_{\mathrm{bol}}/L_{\mathrm{Edd}}$ is inferred from a combination of the $L_{5100\AA}$-$L_{\mathrm H\alpha}$
relation\footnote{The luminosity relation has a rms
scatter around the best-fit line of 0.2dex.} (Greene \& Ho 2005)
\begin{equation}
 L_{5100\AA}=2.4\times10^{43}\bigg(\frac{L_{\mathrm{H\alpha}}}{10^{42}\ \mathrm{ergs\ s^{-1}}}\bigg)^{0.86}\ \mathrm{ergs\ s^{-1}}
\end{equation}
and the bolometric correction of $L_{\mathrm{bol}}=9\lambda L_\lambda(5100\AA)$ (Kaspi et al. 2000).

Because of its bad observed H$\alpha$ line profile,
the parameters $M_{\mathrm{BH}}$ and $L_{\mathrm{bol}}/L_{\mathrm{Edd}}$ of SDSS\,J015950.24+002340.8 are estimated from its broad H$\beta$ emission 
based on a combination of the calibrations of (Vestergaard \& Peterson 2006; Greene \& Ho 2005): 
\begin{equation}
  M_{\mathrm{BH}}=10^{6.67}\bigg(\frac{L_{\mathrm{H\beta}}}{10^{42}\ \mathrm{ergs\ s^{-1}}}\bigg)^{0.63}
\bigg[\frac{\mathrm{FWHM(H\beta)}}{10^3\ \mathrm{km\ s^{-1}}}\bigg]^2 M_\odot
\end{equation}
and 
\begin{equation}
 L_{5100\AA}=7.31\times10^{43}\bigg(\frac{L_{\mathrm{H\beta}}}{10^{42}\ \mathrm{ergs\ s^{-1}}}\bigg)^{0.883}\ \mathrm{ergs\ s^{-1}}
\end{equation}
where the luminosity of its broad H$\beta$ emission $L_{\mathrm{H\beta}}$ is corrected for the local extinction estimated from a combination of 
a Balmer decrement for standard case B recombination and a Galactic extinction curve with $R_V=3.1$. 
Its bolometric luminosity is again obtained from $L_{5100\AA}$ through the bolometric correction of 
$L_{\mathrm{bol}}=9\lambda L_\lambda(5100\AA)$ (Kaspi et al. 2000).

\subsubsection{Statistics}

With the estimated $M_{\mathrm{BH}}$ and $L/L_{\mathrm{Edd}}$, Figure 9 illustrates the 
role of the two parameters on the identified $\Gamma_{\mathrm{2-10keV}}$ versus [OIII]$\lambda$5007 line profile 
correlations.  The related correlation coefficient matrix, which is again based on the 
Spearman rank-order test, is listed in Table 2. On the one hand, the statistics shows that 
both $\Gamma_{\mathrm{2-10keV}}$
and $\delta\upsilon$ are strongly correlated with the estimated $L/L_{\mathrm{Edd}}$.  
The Spearman rank-order tests indicate that the null probabilities of both correlation are smaller than 0.05, which corresponds to 
a significance level larger than 2$\sigma$. Specifically speaking, the significance levels are estimated to be 
$>3.89\sigma$ and 2.89$\sigma$ for the $L/L_{\mathrm{Edd}}$-$\Gamma_{\mathrm{2-10keV}}$ and $L/L_{\mathrm{Edd}}$-$\delta\upsilon$ correlations, respectively.  
For the $\Gamma_{\mathrm{2-10keV}}$ versus $\delta\upsilon$  correlation,
AGNs associated with high $L/L_{\mathrm{Edd}}$ tend to occupy the soft X-ray spectrum end with strong 
[\ion{O}{3}] blue asymmetry, and ones with
low $L/L_{\mathrm{Edd}}$ the hard X-ray spectrum end with weak [\ion{O}{3}] blue asymmetry.
This tendency  implies that $L/L_{\mathrm{Edd}}$ is
a potential physical driver of the $\Gamma_{\mathrm{2-10keV}}$ versus $\delta\upsilon$ correlation. 
On the other hand, in the current sample, $\Delta\upsilon$ is found to be much better correlated with 
$M_{\mathrm{BH}}$ than $L/L_{\mathrm{Edd}}$. 

\section{DISCUSSION}

In this paper, we study the origin of AGN's outflow in its NLR by focusing on the relationship between [\ion{O}{3}]$\lambda$5007 line profile
and hard X-ray emission from the central SMBH. A joint spectral analysis in both optics and hard X-ray allows 
us to reveal a moderate correlation between hard X-ray spectral photon index and [\ion{O}{3}] line asymmetry 
in a sample of 47 local ($z<0.2$) hard X-ray selected type-I AGNs at a significance level of $2.78\sigma$. 
It is noted that the results and implications presented here are
only relevant for the AGNs that are most luminous in the local Universe.

\subsection{$\Gamma_{\mathrm{2-10keV}}-\delta\upsilon$ Correlation: A Connection between Accretion Disk and Outflow in NLR}

\subsubsection{$\Gamma_{\mathrm{2-10keV}}$: an assessment of SMBH accretion}

We argue that the identified hard X-ray spectral photon index versus [\ion{O}{3}]$\lambda5007$ line asymmetry correlation 
(i.e., $\Gamma_{\mathrm{2-10keV}}-\delta\upsilon$ correlation) provides moderate
evidence that the commonly observed AGN's outflow in its NLR (at a radial distance of order 0.1 to 1kpc from the 
central SMBH, e.g., Osterbrock \& Ferland 2006; Heckamn \& Best 2014)  
is related with the accretion process occurring around the central SMBH (i.e., at a distance scale of $\sim10^{1-2}R_s$ from 
the central SMBH, 
where $R_s=2GM_{\mathrm{BH}}/c^2$ is the Schwarzschild radius). 

The parameter $\Gamma_{\mathrm{2-10keV}}$ is believed to be closely linked with the accretion process around central SMBH. 
It is generally believed that the hard X-ray emission of AGN is produced in the region very close to the SMBH.
A commonly accepted scenario of the hard X-ray emission is the accretion disk-corona
model in which a fraction of soft photons from the cold accretion disk is transformed to hard X-ray band through  
the inverse Compton scattering of the hot electrons with a temperature of $\sim10^9$K. These electrons are likely 
accelerated in the corona above the disk by the reconnection of the magnetic fields
(e.g., Haardt \& Maraschi 1991, 1993; Svensson \& Zdziarski 1994; Kawaguchi et al. 2001; Liu et al. 2002, 2003; Cao 2009).  
This model can successfully explain the observed $L/L_{\mathrm{Edd}}-\Gamma$ dependence 
(e.g., Grupe 2004; Desroches et al.
2009; Gierlinski \& Done 2004; Lu \& Yu 1999; Porquet et al.
2004; Wang et al. 2004; Bian 2005, Shemmer et al. 2006, 2008;
Risaliti et al. 2009; Jin et al. 2012; Zhou \& Zhao 2010) as follows.
The scattering can efficiently cool the 
corona above the accretion disk when the disk flux irradiating the corona increases, 
which finally results in a soft, steep X-ray spectrum at high $L/L_{\mathrm{Edd}}$
state (e.g., Pounds et al. 1995).

\subsubsection{$L/L_{\mathrm{Edd}}$ as a physical driver}

We further argue that $L/L_{\mathrm{Edd}}$ is a potential physical 
driver of the $\Gamma_{\mathrm{2-10keV}}-\delta\upsilon$ correlation.  In the current  X-ray type-I
AGN sample, the dependence of $\Gamma_{\mathrm{2-10keV}}$ on 
$L/L_{\mathrm{Edd}}$ can be indirectly learned from the fact that $\Gamma_{\mathrm{2-10keV}}$ is 
found to be correlated with both RFe and FWHM of H$\beta$ (see Table 2) that are basic parameters defining the 
EI space that is widely believed to be physically driven by  $L/L_{\mathrm{Edd}}$.
A direct relationship between 
$\Gamma_{\mathrm{2-10keV}}$ and $L/L_{\mathrm{Edd}}$ can be further identified in the current sample from Table 4 and the bottom panel in Figure 10.

If the above discussion on the physical driver of $\Gamma_{\mathrm{2-10keV}}-\delta\upsilon$ correlation is correct,
a dependence of $\delta\upsilon$ on $L/L_{\mathrm{Edd}}$ (or EI space) is expected in the current sample. 
As shown in Table 2 and the top panel of Figure 10 (9), this dependence at a significance level of 2.89$\sigma$ can be virtually identified in the 
current sample. The dependence is not hard to be understood because it is generally 
believed that high $L/L_{\mathrm{Edd}}$ favors to drive accretion disc winds (e.g., Proga \& Kallman 2002). 
The line strength from BLR is determined by the vertical structure of the accretion disk, governed by 
$L/L_{\mathrm{Edd}}$, in a way in which a large $L/L_{\mathrm{Edd}}$ results in a large X-ray-heated volume that 
generates strong \ion{Fe}{2} complex emission (e.g., BG92).   
In fact, a correlation between [\ion{O}{3}] line profile asymmetry and $L/L_{\mathrm{Edd}}$ 
has been frequently reported in previous studies (e.g., Wang et al. 2011; Wang 2015; Zhang et al. 2011; Boroson 2005; Bian et al. 2005). 
A deficit of extended emission-line region in the AGNs with high $L/L_{\mathrm{Edd}}$ is revealed by
Matsuoka (2012). The deficit could be explained either by the AGN's outflow that blows the gas around central SMBH away
or by galaxy minor merger that produces radio-loud AGNs that are usually associated with an inefficient accretion.

\subsubsection{Disc wind scenario}

So far, two major types of feedback have been proposed for AGNs (Fabian 2012). The one is known as 
the wind/radiation (quasar) mode (e.g., Crenshaw et al. 2003; Pounds
et al. 2003; Ganguly et al. 2007; Reeves et al. 2009; Dunn et al.
2010; King \& Pounds 2003; Murray et al. 1995; King 2003, 2005; Alexander
et al. 2010; King et al. 2011; Zubovas \& King 2012; Proga et al. 2000), and the other is the kinetic (radio) mode 
(e.g., Morganti et al. 2005, 2007; Rosario et al. 2010; Holt et al. 2011; Mahony et al. 2013).       
As recently reviewed in Fabian (2012), there is a current consensus that the two types of feedback 
occur in different AGN types. The wind/radiation mode dominantly operates in luminous quasar phase, while 
the kinetic mode in the less luminous AGNs with low $L/L_{\mathrm{Edd}}$.

The linkage between SMBH accretion disk and outflow in NLR that is diagnosed by the [\ion{O}{3}] line blue wing 
suggests that the observed feedback in NLR 
is rationally originated from the disc winds\footnote{An exclusion of the scenario involving the interaction between radio jet
and interstellar medium  can be found in Section 5.2 for the current sample.}.  A likely scenario is that 
the blue wing of the [\ion{O}{3}] line is likely produced in the inner NLR region (e.g.,
Bian et al. 2005; Wang et al. 2005) in which the kinematics of the emitting gas is dominated by an
acceleration caused by the wind/radiation pressure. 
The disc wind model has been successfully applied to 
explain the observed broad ultraviolet absorption lines in a fraction of $\sim20\%$ quasars and 
the ultra-fast outflows identified from the blue-shifted X-ray \ion{Fe}{25} and \ion{Fe}{26} absorptions in a 
few local AGNs (e.g., Tombesi et al. 2012; Higginbottom et al. 2014). 
In the wind/radiation mode,
a wind can be launched from the inner accretion disk where the ultraviolet photons are emitted from (e.g., 
Murray et al. 1995). The hydrodynamic outflow model calculated by Proga et al. (2008) indicates that the wind
launched from the accretion disk can extended into the inner NLR,   
although the specific launch mechanism is still under debate. Possible mechanisms 
include radiation/line-driven (e.g. Proga et al. 1998, 2000; Laor \& Brandt 2002; Proga \& Kallman 2004; 
Nomura et al. 2013; Higginbottom et al. 2014; Hagino et al.
2015), thermally driven (e.g., Begelman et al. 1983; Krolik \& Kriss 2000), 
magnetically driven (e.g., Blandford \& Payne 1982; Ferreira 1997; Fukumura et al. 2014; Stepanovs \& Fendt 2014) and 
hybrid models (e.g., Everett 2005; Proga 2003).

\subsection{Radio Emission and [OIII] Line Profile}

Some previous studies argued that the outflow in NLR is potentially driven by the interaction between radio jet and interstellar
medium (e.g., Heckman et al. 1984; Whittle 1985; Brotherton 1996; Holt et al. 2008, 2011; Nesvadba et al. 2008; Whittle \& Wilson 2004;
Guillard 2012; Morganti et al. 2007; Mahony et al. 2013). This argument is recently reinforced by the studies in 
Mullaney et al. (2013) and Zakamska \& Greene (2014). Both study claim a relation between [\ion{O}{3}] line asymmetry and 
radio luminosity in both type I and II AGNs observed by SDSS.  

The role of radio emission played in the [\ion{O}{3}] line profile is test for the 47 X-ray selected type I AGNs used in this 
study. The 47 X-ray AGNs are cross-matched with the FIRST survey catalog (Becker et al. 2003). 
This environmental cross-match returns only 20 radio objects with detected radio flux exceeding the FIRST
limiting flux density (5$\sigma$) of 1mJy. The upper limit of radio flux is also taken from the FIRST survey
for the each of the other 27 sources, basing upon the reported detection limit at the corresponding celestial position.  
The luminosity at 1.4 GHz (rest frame) of each radio source at a given redshift $z$ is calculated from the observed integrated
flux density at 1.4 GHz $f_\nu$ through $L_{\mathrm{1.4GHz}}=4\pi d_L^2 f_\nu(1+z)^{-1-\alpha}$, 
where $d_L$ is the luminosity distance, and $\alpha=-0.8$
(e.g., Ker et al., 2012) is the spectral slope defined as $f_\nu\propto\nu^{\alpha}$. The calculated
$L_{\mathrm{1.4GHz}}$ ranges from $10^{22}$ to $10^{26}\ \mathrm{W\ Hz^{-1}}$, and 
is plotted against $\Gamma_{\mathrm{2-10keV}}$, $\delta\upsilon$ and $\Delta\upsilon$ in the left, middle and right panel in Figure 11, respectively.
The solid blue points denote the sources with a detected radio flux, and the open red points the ones with 
a flux upper limit. 
The corresponding statistics based on Kendall's $\tau$ is tabulated in Table 2. The values tabulated in 
line (6) are based on the 20 sources with a detected radio flux, and the ones in line (7) on all the 47 sources through 
the survival analysis with non-parametric model (e.g., Isobe et al. 1986). 
Our statistics shows that in the current sample there is no evidence that both blue wing and bulk velocity blueshift of the 
[\ion{O}{3}] line are driven by the radio emission, i.e., no direct relationship can be identified between 
$L_{\mathrm{1.4GHz}}$ and $\delta\upsilon$ ($\Delta\upsilon$) in the current sample. 

We argue that this result does not mean an inevitable disagreement on the previous studies.  
At first, one should be bear in mind that the relationships related with radio luminosity are hard to be 
firmly tested in the current study because its sample size (47 sources) is significantly less than those (with hundreds to thousands of sources) in
Mullaney et al. (2013) and Zakamska \& Greene (2014). Secondly, the middle panel in Figure 11
shows that the most significant [\ion{O}{3}] blue asymmetry tends to occur in the objects with a radio luminosity at 1.4GHz of 
$10^{22}-10^{24} \mathrm{W\ Hz^{-1}}$, which is close to that observed in Mullaney et al. (2013). 
Zakamska \& Greene (2014) recently proposed that the radio emission in radio-quite AGNs might be produced by the 
accelerated particles in the interstellar medium of the host galaxy that is shocked by the accretion disk wind.

\subsection{Evolution of Feedback in AGNs}

We close the paper by a short discussion on the issue of co-evolution of AGN's feedback and its host galaxy. 
We argue that the revealed X-ray emission (and $L/L_{\mathrm{Edd}}$) dependent outflow seen in NLR is consistent 
with the coevolution scenario that was suggested in many previous studies. In fact, both AGN's X-ray 
emission (and $L/L_{\mathrm{Edd}}$) and [\ion{O}{3}] line profile have been claimed to be related with the 
host galaxy stellar population age. On the one hand, Wang et al. (2013) identified a correlation between AGN's hard X-ray 
spectral index and host galaxy stellar population age in X-ray selected SDSS type-II AGNs: harder the 
X-ray spectrum, older the host stellar population will be. The important role of $L/L_{\mathrm{Edd}}$ 
in the coevolution issue has been frequently revealed in previous studies by studying the relationship 
between $L/L_{\mathrm{Edd}}$ and host stellar population (e.g., Heckman \& Kauffmann 2006;
Goulding et al. 2010; Kewley et al. 2006; Wang et al. 2006; Wang \& Wei 2008, 2010; Kauffmann et al. 2007; Wild et al. 2007; 
Wang 2015). On the other hand, by analyzing the optical spectra of narrow emission-line galaxies taken from SDSS survey,
Wang et al. (2011) proposed a trend in which 
AGNs with stronger blue asymmetries tend to be associated with younger stellar populations. 
This result is recently confirmed and reinforced in Wang (2015) by focusing on partially obscured AGNs. 
Combining these results implies that the SMBH growth through gas accretion and host galaxy building is potentially
linked by the outflow launched from the accretion disk. 
It is generally believed that
a self-regulated SMBH growth and host star formation can be produced by suppressing the star formation
in both galaxies merger and secular evolution scenarios through 
the feedback from central AGN that sweeps out circumnuclear gas (e.g., Kormendy \& Ho 2012; Alexander \& Hickox 2012; Fabian 2012; 
Zubovas et al. 2013).

The dependence of outflow on both SMBH accretion properties and host stellar population suggests that 
the feedback process, not as a constant, likely evolves with the SMBH growth, and hence with host star formation in an AGN recycle.
That means a strong feedback is required to regulate SMBH mass growth and host star formation in the early gas-rich phase 
associated with soft X-ray spectrum, high $L/L_{\mathrm{Edd}}$ and young stellar population. 
The regulated SMBH growth and host star formation can be, however, achieved by a weak feedback in the late gas-poor phase when 
both accretion and starforming activities become to be weak. 


\section{CONCLUSION}

We study the origin of AGN's outflow occurring in its NLR by focusing on the relationship between [\ion{O}{3}]$\lambda$5007 line profile
and hard X-ray emission from the central SMBH  in a sample of 47  local X-ray selected type I AGNs ($z<0.2$). 
These luminous AGNs are extracted from the 2XMMi/SDSS DR7 catalog, and have X-ray luminosities in 2-10keV in a 
range from $10^{42}$ to $10^{45}\ \mathrm{erg\ s^{-1}}$. 
A joint spectral analysis in both optics and hard X-ray on the sample allows us to identify a moderate  correlation with 
a significance level of 2.78$\sigma$,
in which luminous AGNs with more significant [\ion{O}{3}] blue asymmetry tend to be associated with 
steeper X-ray spectra. Our statistics show that the correlation is related with 
$L/L_{\mathrm{Edd}}$ at a 2-3$\sigma$ significance level, which suggests that the AGN's outflow in 
its NLR is likely driven by the accretion process occurring around the central SMBH.

\acknowledgments


The authors thank the anonymous referee for his/her careful review and helpful suggestions
for improving the manuscript. We thanks Dr. X. L. Zhou for the help in X-ray spectral analysis.
This study uses the
SDSS archive data that was created and distributed by the Alfred P. Sloan Foundation.
This work is based on observations obtained with XMM-Newton, an ESA science mission with
instruments and contributions directly funded by ESA Member
States and the USA (NASA).
The study is supported by the National Basic Research Program of China (grant
2009CB824800) and by National Natural Science
Foundation of China under grants 11473036 and 11273027.

\clearpage

\begin{deluxetable}{llccccccccc}
\tabletypesize{\scriptsize}
\rotate
\tablecaption{Properties of the XMM-Newton/SDSS-DR7 type-I AGNs\label{tbl-2}}
\tablewidth{0pt}
\tablehead{
\colhead{SDSS} & \colhead{z} & \colhead{$\mathrm{FWHM_{H\beta}}$} &  \colhead{$\log \frac{L_{\mathrm{H\alpha}}}{\mathrm{erg\ s^{-1}}}$} & \colhead{RFe} & 
\colhead{$\Gamma_{\mathrm{2-10keV}}$} & \colhead{$\Delta\upsilon$\tablenotemark{b}} & \colhead{$\log \frac{M_{\mathrm{BH}}}{M_\odot}$} & 
\colhead{$L/L_{\mathrm{Edd}}$} & \colhead{$\delta\upsilon$}  & \colhead{$\log \frac{L_{\mathrm X}}{\mathrm{erg\ s^{-1}}}$}\\
\colhead{} & \colhead{} & \colhead{$\mathrm{km\ s^{-1}}$} & 
\colhead{} & \colhead{} & \colhead{} & \colhead{$\mathrm{km\ s^{-1}}$} & \colhead{} & \colhead{} & \colhead{$\mathrm{km\ s^{-1}}$} & \colhead{} \\
\colhead{(1)} & \colhead{(2)} & \colhead{(3)} & \colhead{(4)} & \colhead{(5)} & \colhead{(6)} & \colhead{(7)} & 
\colhead{(8)} & \colhead{(9)} & \colhead{(10)} & \colhead{(11)} }

\startdata
J010712.04+140844.9     &  0.0769 &  $1420\pm220$ &  41.3 &   $0.55\pm0.09$ &  $2.15_{-0.10}^{+0.10}$ &   $-25.1\pm39.7$ &  6.19 &  0.33     &   $-39.1\pm13.6$          &  42.5  \\
J015950.24+002340.8    &  0.1627 &  $3240\pm150$ &  42.7\tablenotemark{a} &   $0.57\pm0.03$ &  $1.96_{-0.19}^{+0.17}$ &    $56.3\pm25.0$ &  8.09 &  0.19     &   $-332.1\pm185.5$        &  43.7  \\
J030639.58+000343.2     &  0.1074 &  $1970\pm50$  &  43.7 &   $0.003\pm0.002$ &  $1.80_{-0.04}^{+0.04}$ &   $108.3\pm34.9$  &  8.13 &  0.38    &   $-85.0\pm40.9$        &  43.4  \\
J091848.61+211717.0     &  0.1493 &  $1740\pm190$ &  43.8 &   $0.05\pm0.01$ &  $2.04_{-0.19}^{+0.19}$ &   $103.7\pm45.9$ &  7.80 &  1.03     &   $-212.8\pm37.3$         &  42.9  \\
J092247.02+512038.0     &  0.1598 &  $2610\pm150$ &  42.3 &   $2.46\pm0.23$ &  $2.23_{-0.10}^{+0.10}$ &  $-361.2\pm45.1$ &  6.92 &  0.44     &   $-349.1\pm169.6$        &  43.4  \\
J092343.00+225432.6     &  0.0332 &  $2750\pm100$ &  41.5 &   $0.25\pm0.02$ &  $1.90_{-0.02}^{+0.02}$ &   $-42.6\pm17.6$ &  6.86 &  0.09     &   $-42.4\pm29.6$          &  43.6  \\
J093922.89+370943.9     &  0.1859 &  $1900\pm150$ &  42.6 &   $0.61\pm0.05$ &  $2.10_{-0.39}^{+0.35}$ &   $-29.0\pm42.7$ &  7.13 &  0.50     &   $-424.7\pm362.5$        &  43.2  \\
J094439.88+034940.1     &  0.1554 &  $4410\pm30$  &  43.0 &   $0.49\pm0.06$ &  $1.90_{-0.18}^{+0.19}$ &   $-31.4\pm68.6$  &  8.24 &  0.08     &   $-165.3\pm108.6$       &  43.2  \\
J100035.47+052428.5*    &  0.0786 &  $4020\pm130$ &  41.8 &   $0.06\pm0.05$ &  $1.41_{-0.22}^{+0.22}$ &    $22.7\pm8.9$  &  6.78 &  0.20     &   $-7.0\pm3.9$            &  43.2  \\
J102822.84+235125.7*    &  0.1734 &  $5340\pm200$ &  42.5 &   $0.01\pm0.01$ &  $1.89_{-0.20}^{+0.19}$ &    $98.2\pm18.8$ &  8.05 &  0.04     &   $+5.5\pm23.8$           &  43.4  \\
J103059.09+310255.7*    &  0.1781 &  $5560\pm330$ &  43.5 &   $0.00\pm0.00$ &  $1.59_{-0.07}^{+0.07}$ &    $-0.6\pm27.5$ &  8.71 &  0.07     &   $+22.2\pm12.9$       &  44.3  \\
J103349.93+631830.4     &  0.1555 &  $1950\pm410$ &  42.8 &   $0.03\pm0.01$ &  $2.11_{-0.13}^{+0.14}$ &   \dotfill        &  7.48 &  0.28    &   $-42.9\pm40.8$          &  43.1  \\
J103438.59+393828.2     &  0.0431 &  $ 870\pm90 $ &  42.1 &   $0.16\pm0.02$ &  $2.56_{-0.17}^{+0.17}$ &   $-27.3\pm15.7$  &  6.61 &  0.53     &   $-191.0\pm32.9$      &  42.3  \\
J105143.89+335926.7     &  0.1671 &  $3880\pm160$ &  43.3 &   $0.18\pm0.01$ &  $1.92_{-0.04}^{+0.04}$ &   $ 75.2\pm41.0$ &  8.15 &  0.17     &   $-1.1\pm5.7$    &  44.0  \\
J110101.77+110248.9*    &  0.0356 &  $7210\pm250$ &  41.9 &   $0.01\pm0.01$ &  $1.66_{-0.03}^{+0.05}$ &     $0.1\pm16.1$ &  7.79 &  0.02     &   $+5.5\pm8.0$    &  43.0  \\
J111706.39+441333.3     &  0.1438 &  $5180\pm170$ &  44.2 &   $0.02\pm0.01$ &  $1.52_{-0.02}^{+0.02}$ &   $-12.6\pm49.0$ &  8.88 &  0.21     &   $-124.7\pm83.5$      &  44.0  \\
J111830.28+402554.0     &  0.1545 &  $2100\pm110$ &  43.6 &   $0.43\pm0.02$ &  $2.14_{-0.11}^{+0.10}$ &  $-109.8\pm40.4$ &  7.70 &  0.54     &   $-116.1\pm48.2$      &  43.8  \\
J112328.11+052823.2     &  0.1013 &  $1660\pm350$ &  42.2 &   $0.18\pm0.04$ &  $2.00_{-0.11}^{+0.11}$ &    $11.5\pm35.4$  &  7.04 &  0.25     &   $-71.6\pm7.7$          &  42.8  \\
J114008.71+030711.4     &  0.0811 &  $1350\pm80$  &  41.5 &   $0.78\pm0.07$ &  $1.84_{-0.28}^{+0.27}$ &    $-3.7\pm22.9$  &  6.52 &  0.22     &   $-62.0\pm34.4$         &  42.6  \\
J120442.10+275411.7*    &  0.1651 &  $5080\pm420$ &  43.3 &   $0.06\pm0.02$ &  $1.58_{-0.08}^{+0.07}$ &    $38.4\pm9.4$  &  8.28 &  0.12     &   $-232.0\pm27.7$         &  44.4  \\
J121356.19+140431.3     &  0.1539 &  $4970\pm460$ &  42.6 &   $0.15\pm0.01$ &  $1.48_{-0.20}^{+0.18}$ &   \dotfill       &  8.20 &  0.04     &   $-53.7\pm113.3$         &  43.4  \\
J121930.87+064334.4     &  0.0804 &  $1780\pm70$  &  43.0 &   $0.08\pm0.01$ &  $2.11_{-0.08}^{+0.08}$ &    $11.5\pm43.9$  &  7.42 &  0.54     &   $-61.0\pm22.5$       &  43.0  \\
J122137.93+043026.1*    &  0.0947 &  $8530\pm410$ &  42.0 &   $0.25\pm0.01$ &  $1.43_{-0.19}^{+0.17}$ &    $21.6\pm15.9$ &  8.18 &  0.01     &   $+2.1\pm4.3$    &  42.9  \\
J123113.66+151127.9     &  0.1919 &  $3500\pm190$ &  42.3 &   $0.47\pm0.03$ &  $2.20_{-0.15}^{+0.16}$ &    $17.9\pm44.1$ &  7.48 &  0.12     &   $-83.7\pm48.5$          &  43.3  \\
J124013.80+473354.8     &  0.1174 &  $1840\pm240$ &  41.7 &   $0.75\pm0.06$ &  $1.48_{-0.48}^{+0.45}$ &     $4.7\pm51.2$ &  6.72 &  0.20     &   $-33.1\pm10.7$          &  43.0  \\
J124210.60+331702.6     &  0.0437 &  $1470\pm120$ &  43.0 &   $0.06\pm0.01$ &  $2.15_{-0.01}^{+0.01}$ &    $-2.9\pm35.1$ &  7.32 &  0.60     &   $-173.1\pm44.5$         &  43.3  \\
J124635.24+022208.7     &  0.0482 &  $1270\pm90$  &  41.9 &   $0.45\pm0.05$ &  $2.36_{-0.10}^{+0.10}$ &  $-118.4\pm32.8$  &  6.72 &  0.31     &   $-62.3\pm43.8$         &  43.0  \\
J130022.15+282402.6     &  0.0911 &  $3550\pm140$ &  42.3 &   $0.39\pm0.01$ &  $1.70_{-0.12}^{+0.11}$ &     $7.5\pm48.7$ &  7.46 &  0.12     &   $-7.9\pm8.2$            &  43.4  \\
J130947.00+081948.2     &  0.1543 &  $4320\pm130$ &  43.7 &   $0.10\pm0.01$ &  $1.46_{-0.04}^{+0.04}$ &    $41.4\pm24.6$ &  8.43 &  0.20     &   $-137.1\pm25.9$      &  44.0  \\
J133141.02-015212.4     &  0.1454 &  $1570\pm280$ &  42.1 &   $0.24\pm0.05$ &  $1.72_{-0.41}^{+0.39}$ &   $-38.7\pm31.0$ &  7.01 &  0.22     &   $-27.4\pm26.8$       &  43.0  \\
J134351.06+000434.7*    &  0.0737 &  $3300\pm150$ &  41.7 &   $0.69\pm0.01$ &  $1.43_{-0.23}^{+0.23}$ &    $18.3\pm30.8$ &  6.84 &  0.16     &   $+0.9\pm4.5$    &  41.6  \\
J134834.94+263109.8*    &  0.0589 &  $1580\pm90$  &  42.3 &   $0.11\pm0.01$ &  $1.79_{-0.23}^{+0.23}$ &    $11.7\pm37.6$  &  7.11 &  0.28    &   $-119.5\pm56.9$         &  42.7  \\
J135435.68+180517.4     &  0.1509 &  $4190\pm110$ &  43.9 &   $0.04\pm0.01$ &  $1.83_{-0.11}^{+0.11}$ &   \dotfill       &  8.52 &  0.25     &   $-142.5\pm48.9$      &  44.0  \\
J135553.52+383428.7*    &  0.0502 &  $6500\pm220$ &  42.4 &   $0.04\pm0.01$ &  $1.56_{-0.07}^{+0.07}$ &    $22.6\pm7.3$  &  7.75 &  0.07     &   $-25.3\pm7.4$           &  43.2  \\
J135724.51+652505.9     &  0.1063 &  $1590\pm190$ &  41.6 &   $0.30\pm0.05$ &  $1.96_{-0.23}^{+0.18}$ &   $-15.5\pm27.9$ &  6.40 &  0.36     &   $-13.2\pm9.8$           &  42.9  \\
J140251.19+263117.5     &  0.1875 &  $5900\pm80$  &  44.1 &   $0.07\pm0.01$ &  $1.49_{-0.10}^{+0.10}$ &    $104.5\pm33.2$ &  9.01 &  0.11     &   $-84.3\pm26.4$         &  44.2  \\
J140621.89+222346.5     &  0.0979 &  $4280\pm110$ &  42.5 &   $1.43\pm0.04$ &  $1.95_{-0.42}^{+0.44}$ &   $-47.2\pm44.3$ &  7.83 &  0.07     &   $-283.1\pm105.0$     &  42.5  \\
J140700.40+282714.6     &  0.0766 &  $8240\pm450$ &  43.5 &   $0.09\pm0.05$ &  $1.22_{-0.10}^{+0.10}$ &    $61.5\pm20.6$ &  9.18 &  0.03     &   $-29.6\pm15.0$          &  42.5  \\
J141519.50-003021.5     &  0.1347 &  $2260\pm190$ &  41.8 &   $1.21\pm0.11$ &  $2.15_{-0.09}^{+0.09}$ &  $-304.5\pm58.3$ &  6.94 &  0.14     &   $-99.6\pm67.6$          &  42.9  \\
J141700.82+445606.3     &  0.1136 &  $2590\pm150$ &  42.8 &   $0.86\pm0.07$ &  $1.93_{-0.09}^{+0.09}$ &   \dotfill       &  7.73 &  0.18     &   $-276.9\pm49.7$         &  43.5  \\
J143452.45+483942.7     &  0.0365 &  $4520\pm190$ &  43.0 &   $0.004\pm0.003$ &  $1.58_{-0.23}^{+0.21}$ &   $-39.5\pm45.4$ &  8.23 &  0.08    &   $+14.8\pm6.3$        &  43.1  \\
J145108.76+270926.9     &  0.0645 &  $3040\pm100$ &  42.6 &   $0.79\pm0.03$ &  $2.27_{-0.06}^{+0.06}$ &     $9.9\pm12.9$ &  7.50 &  0.20     &   $-103.7\pm10.6$         &  43.3  \\
J150626.44+030659.9     &  0.1734 &  $2270\pm400$ &  42.8 &   $0.15\pm0.01$ &  $1.79_{-0.25}^{+0.02}$ &  \dotfill        &  7.49  &  0.28    &   $-147.7\pm20.8$         &  43.6  \\
J151600.96+000949.7     &  0.1712 &  $4440\pm190$ &  42.3 &   $0.09\pm0.11$ &  $1.68_{-0.07}^{+0.07}$ &   \dotfill       &  7.81 &  0.05     &   $-45.2\pm38.0$          &  43.4  \\
J155909.63+350147.5     &  0.0311 &  $1700\pm110$ &  41.7 &   $1.53\pm0.14$ &  $2.15_{-0.07}^{+0.06}$ &  $ -70.9\pm21.7$ &  6.42 &  0.41     &   $-18.7\pm23.2$       &  42.9  \\
J160452.45+240241.6*    &  0.0876 &  $2710\pm780$ &  42.7 &   $0.17\pm0.04$ &  $1.78_{-0.07}^{+0.07}$ &    $42.7\pm58.4$ &  7.56 &  0.21     &   $-17.9\pm6.0$   &  43.1  \\
J221918.53+120753.1     &  0.0815 &  $1110\pm80$  &  42.4 &   $0.19\pm0.01$ &  $2.47_{-0.13}^{+0.12}$ &   $-26.3\pm32.6$  &  6.80 &  0.70     &   $-63.7\pm27.2$         &  43.1  \\
\enddata
\tablecomments{The object that shows a Seyfert-1.5 like spectrum is marked by a star. The redshifts given in Column (2) are provided by the SDSS pipelines.}
\tablenotetext{a}{The value is given for H$\beta$ rather than H$\alpha$ because of the bad observed H$\alpha$ line profile. See Section 4.3.1 in the text for the details.}
\tablenotetext{b}{The values of $\Delta\upsilon$ are not available for a few objects since the bad constraint on their H$\beta$ narrow peaks.}
\end{deluxetable}

\clearpage

\begin{table}
\begin{center}
\caption{Correlations coefficient matrix related with the hard X-ray photon index versus [\ion{O}{3}]$\lambda$5007 emission line profile correlations\label{tbl-2}}
\begin{tabular}{lcccc}
\tableline\tableline
Property  & $\Gamma_{\mathrm{2-10keV}}$ & $\delta\upsilon$ & $\Delta\upsilon$\\
(1) & (2) & (3) & (4) \\ 
\tableline
(1) $\Gamma_{\mathrm{2-10keV}}$ & \dotfill &  -0.411(0.0054)  & -0.483(0.0022)\\
(2) RFe\dotfill                  &  0.408(0.0057)    & -0.268(0.0694) & -0.508(0.0011)\\
(3) $\mathrm{FWHM_{H\beta}}$     & -0.673($<10^{-4}$) &  0.269(0.0685) &  0.416(0.0077)\\ 
(4) $M_{\mathrm{BH}}$\dotfill    & -0.497(0.0008)    & -0.073(0.6218) &  0.509(0.0011)\\
(5) $L/L_{\mathrm{Edd}}$\dotfill &  0.620($<10^{-4}$) & -0.418(0.0046) & -0.266(0.0886)\\ 
(6) $L_{\mathrm{1.4GHz}}$\tablenotemark{a}\dotfill& -0.240(0.1510)    & -0.170(0.3103) &   0.281(0.0929)\\
(7) $L_{\mathrm{1.4GHz}}$\tablenotemark{b}\dotfill&  0.010(0.8882)    & -0.107(0.1362) &   0.085(0.2371)\\         
\tableline
\tablenotetext{a}{The statistics is based on the 20 sources with detected radio flux taken from the FIRST catalog.}
\tablenotetext{b}{The statistics is based on all the 47 sources through survival analysis. }
\end{tabular}
\end{center}
\end{table}




\clearpage
\begin{figure}
\epsscale{.80}
\plotone{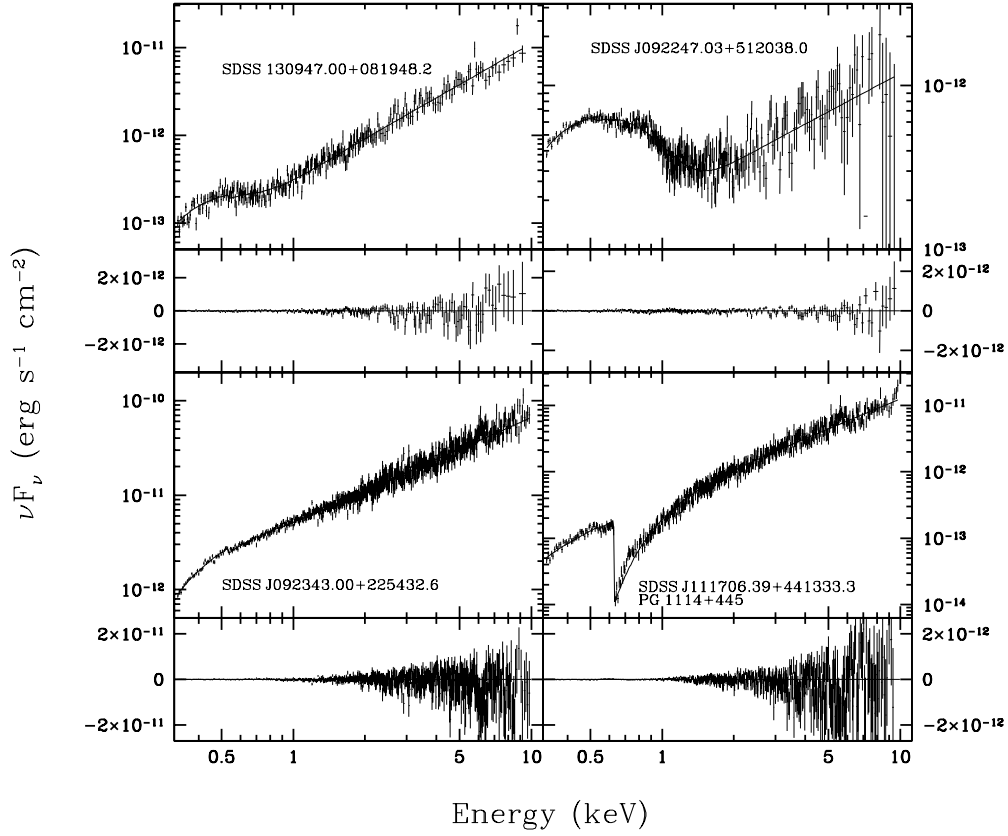}
\caption{Examples of the X-ray spectral modelings. In each panel, the top sub-panel shows the EPIC PN X-ray spectrum
in terms of $\nu F_\nu$  
and the best-fit spectral model (see  Section 3.1 for the details of the used model).  A strong absorption
edge at $0.72\pm0.01\ \mathrm{keV}$ (rest frame) is required in SDSS\,J111706.39+441333.3 (PG\,1114+445, right-bottom panel)
to properly reproduce its observed spectra. The bottom sub-panel shows the 
deviation of the observed data from the best-fit model in terms of $\nu F_\nu$.
}
\end{figure}

\begin{figure}
\epsscale{.80}
\plotone{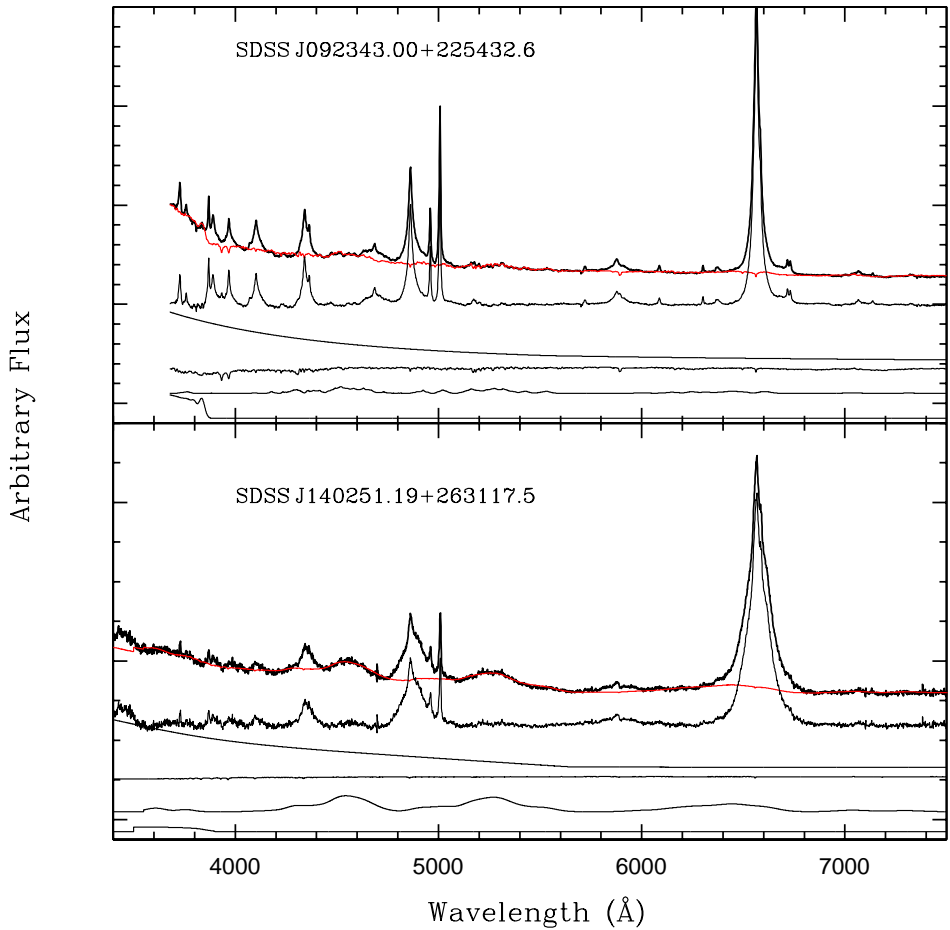}
\caption{
Illustration of the modeling and removal of the continuum for two typical sources.
In each panel, the top curve shows the observed rest-frame spectrum overplotted by 
the modeled continuum by the red curve. The continuum-removed emission-line spectrum 
is shown below the observed one. The modeled continuum is obtained by a reddened linear combination of 
a broken power law from the central AGN, a starlight component from the host galaxy,
the emission from the \ion{Fe}{2} complex, the Balmer continuum, and the high order Balmer emission lines,
which are plotted in ordinals below the emission-line spectrum. The intrinsic extinction is 
considered in the modeling by using a Galactic extinction curve with $R_V=3.1$.      
All the spectra are shifted vertically by an arbitrary amount for visibility.
}
\end{figure}

\begin{figure}
\epsscale{.80}
\plotone{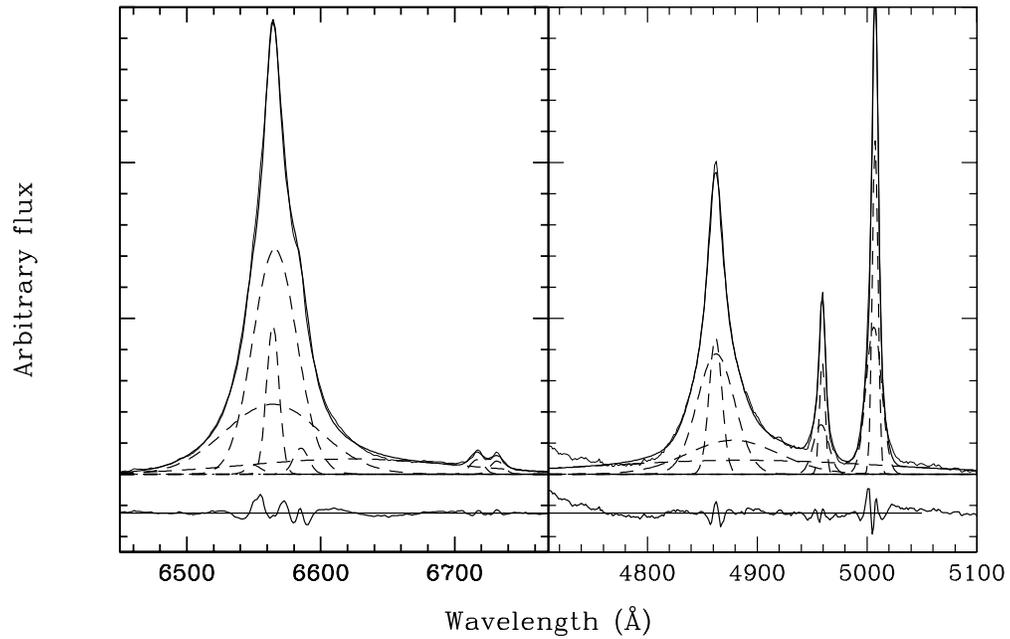}
\caption{
Line profile modelings of SDSS\,J092343.00+225432.6 by a linear combination a set of Gaussian functions
for the H$\alpha$ (left panel) and H$\beta$ (right panel) regions. In each panel,
the observed and modeled line profiles are plotted by light and heavy
solid lines, in which the modeled continuum has already been removed from the original observed spectrum.  
Each Gaussian function is shown by a dashed
line. The sub-panel underneath each line spectrum presents the residuals
between the observed and modeled profiles.
}
\end{figure}

\begin{figure}
\epsscale{.50}
\plotone{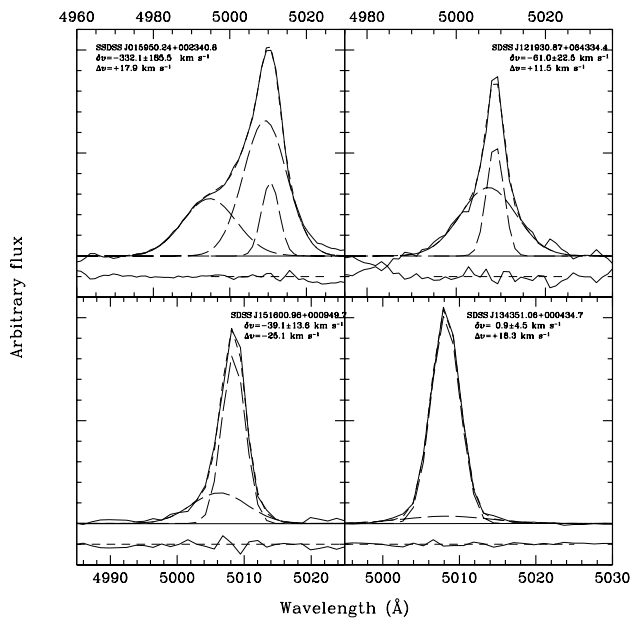}
\caption{[\ion{O}{3}]$\lambda5007$ line profile modeling by a sum of $n$ Gaussian functions
for four typical cases.  In each panel,
the observed and modeled line profiles are plotted by solid and dashed lines, 
respectively. The sub-panel underneath each line spectrum presents the residuals
between the observed and modeled profiles. All the spectra are transformed to rest frame based on the 
redshifts given by the SDSS pipelines.
}
\end{figure}

\begin{figure}
\epsscale{.50}
\plotone{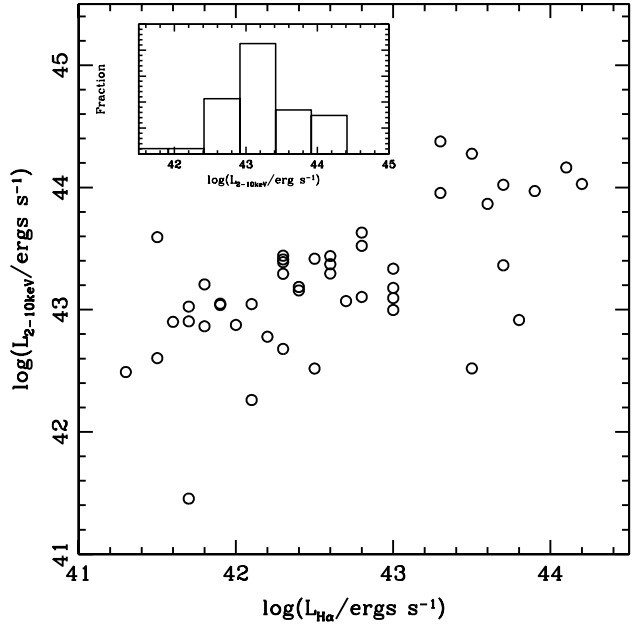}
\caption{Hard X-ray Luminosity in the energy bandpass 2-10 keV ($L_{\mathrm{2-10keV}}$) plotted against the 
luminosity of broad H$\alpha$ emission. 
The inset panel shows the distribution of $L_{\mathrm{2-10keV}}$.}
\end{figure}

\begin{figure}
\epsscale{.50}
\plotone{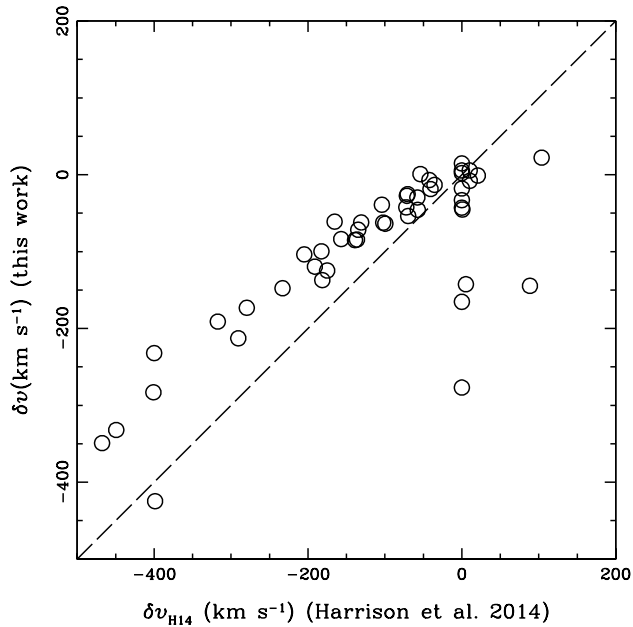}
\caption{A comparison between the value of $\delta\upsilon$ used in this paper and the velocity offset $\delta\upsilon_{\mathrm{H14}}$ defined in 
Harrison et al. (2014). See the text for the details of the definition of $\delta\upsilon_{\mathrm{H14}}$. A ratio of 1 is presented by the dashed line.}
\end{figure}

\begin{figure}
\epsscale{.80}
\plotone{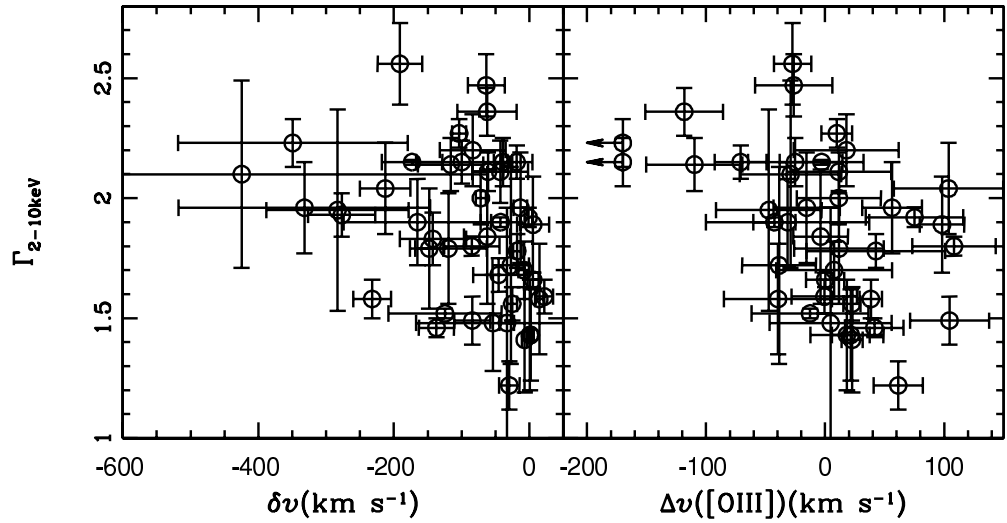}
\caption{\it Left panel: \rm The hard X-ray photon spectral index ($\Gamma_{\mathrm{2-10keV}}$) plotted as a function of
[\ion{O}{3}]$\lambda5007$ line profile asymmetry parameter $\delta\upsilon$. \it Right panel: \rm The same as the left one but for [\ion{O}{3}] line bulk velocity 
shift with respect to the H$\beta$ narrow peak ($\Delta\upsilon$). The two points associated with left arrows mark the objects
with $\Delta\upsilon$ as large as $\sim300\ \mathrm{km\ s^{-1}}$.}
\end{figure}

\begin{figure}
\epsscale{.80}
\plotone{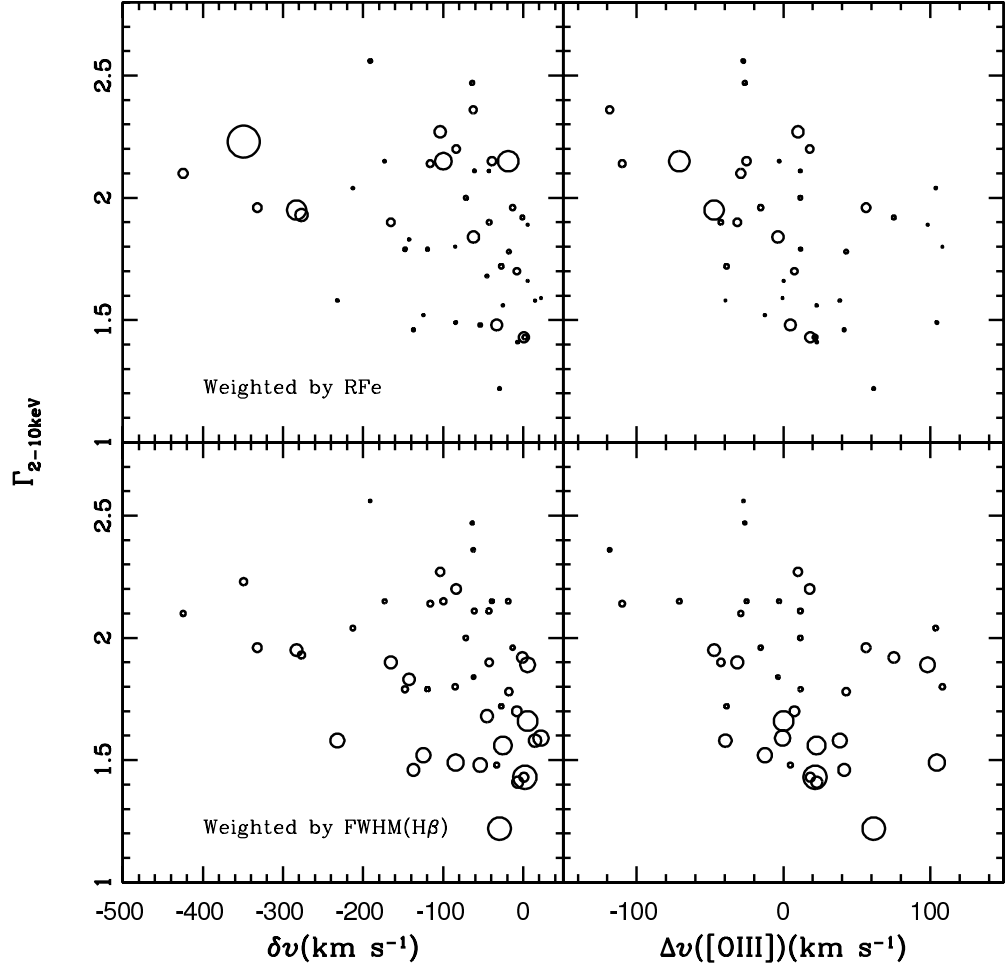}
\caption{\it Bottom panels: \rm The $\Gamma_{\mathrm{2-10keV}}-h_3$ (left) and $\Gamma_{\mathrm{2-10keV}}-\Delta\upsilon$ (right)
correlations in which the size of each point is proportional to the measured FWHM of H$\beta$ broad emission line. \it Top panels: \rm
The same as the bottom ones but for the point size that is proportional to RFe. 
}
\end{figure}

\begin{figure}
\epsscale{.80}
\plotone{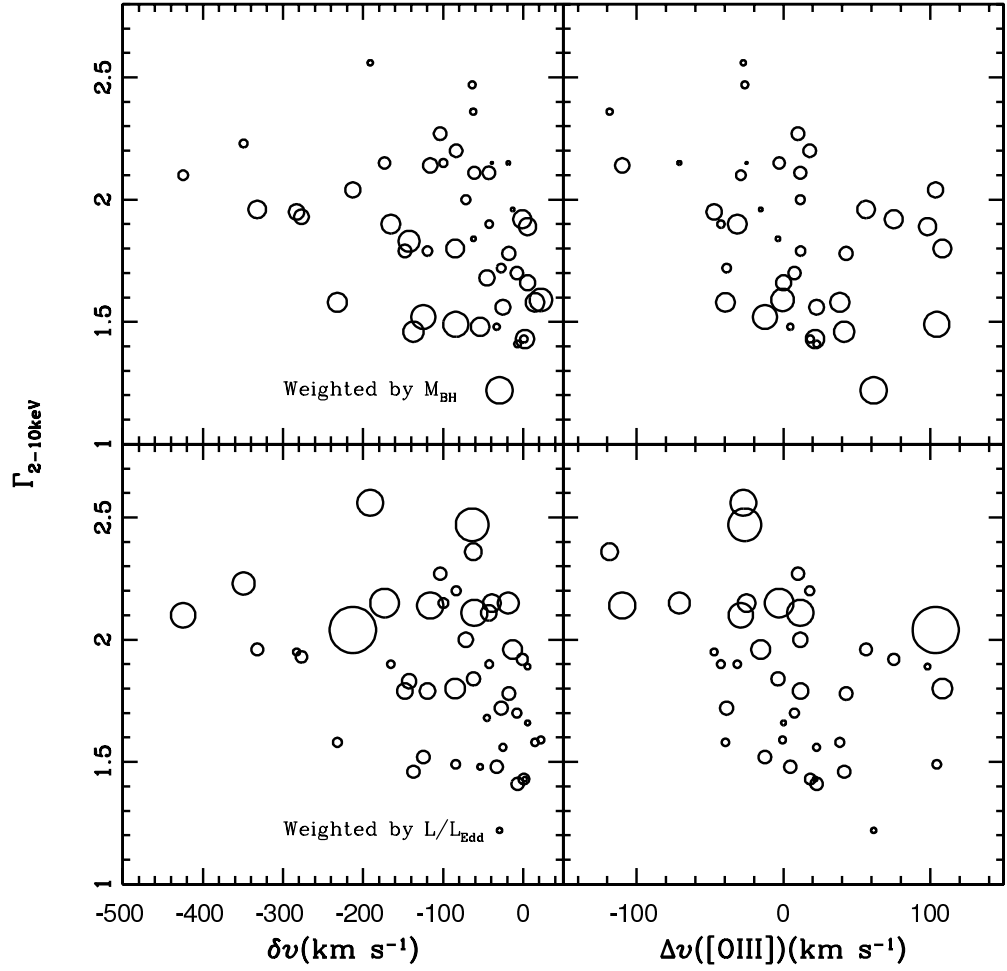}
\caption{\it Bottom panels: \rm The $\Gamma_{\mathrm{2-10keV}}-h_3$ (left) and $\Gamma_{\mathrm{2-10keV}}-\Delta\upsilon$ (right)
correlations in which the size of each point is proportional to the estimated $L/L_{\mathrm{Edd}}$. \it Top panels: \rm
The same as the bottom ones but for the point size that is proportional to  $M_{\mathrm{BH}}$.}
\end{figure}

\begin{figure}
\epsscale{.50}
\plotone{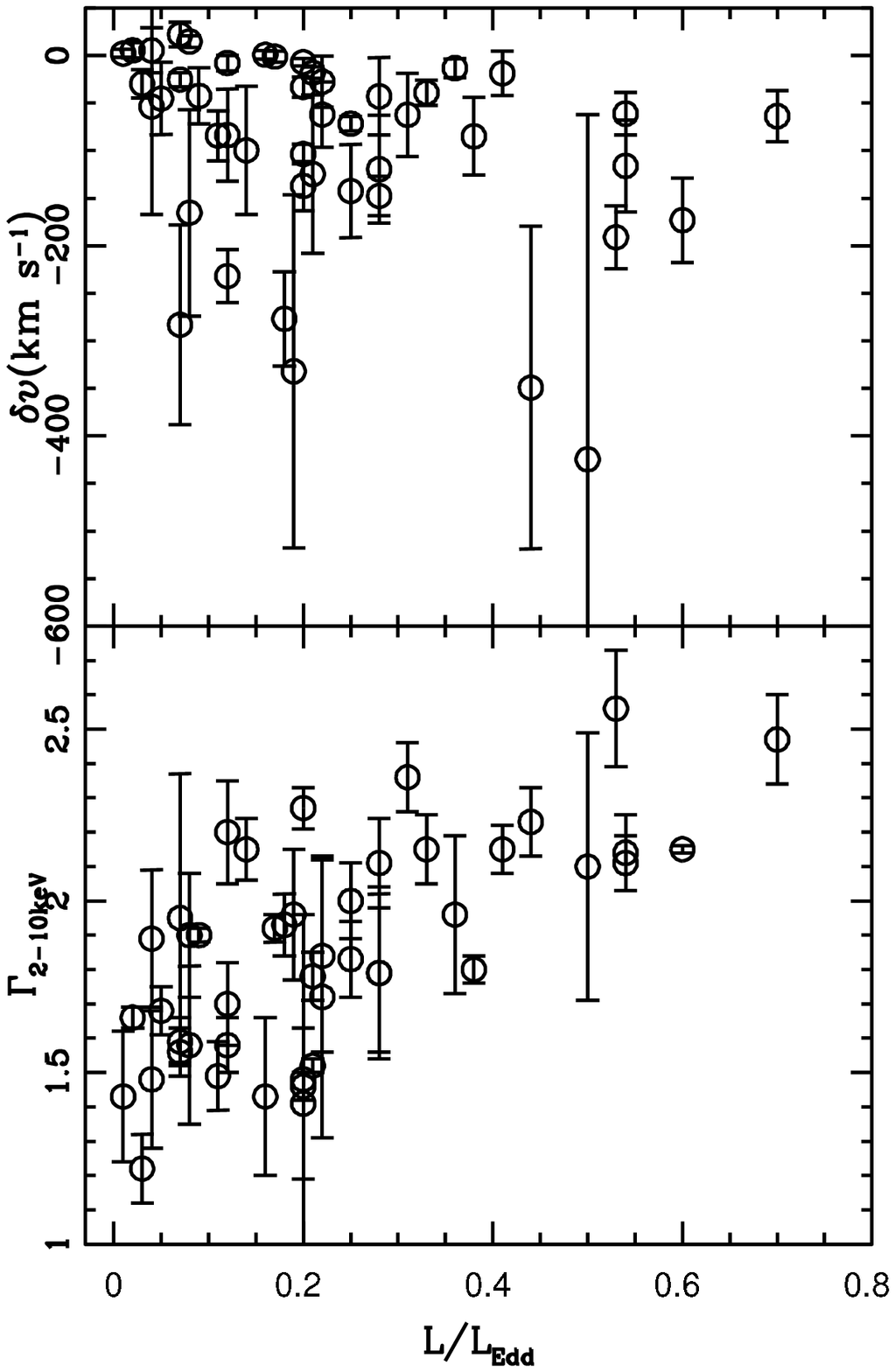}
\caption{$\Gamma_{\mathrm{2-10keV}}$ (\it bottom panel\rm) and $\delta\upsilon$ (\it top panel\rm) are plotted as a function of $L/L_{\mathrm{Edd}}$.} 
\end{figure}

\begin{figure}
\epsscale{.80}
\plotone{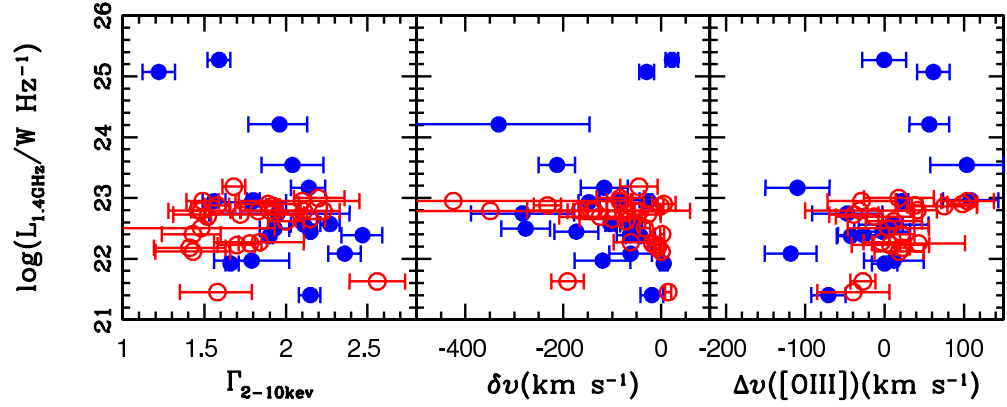}
\caption{Calculated radio luminosity at 1.4GHz (rest frame) plotted against $\Gamma_{\mathrm{2-10keV}}$, $\delta\upsilon$ and $\Delta\upsilon$ in 
the left, middle and right panels, respectively. The sources with a detected radio flux is shown by the blue solid points, and
the ones with a flux upper limit by the red open points. }
\end{figure}

\end{document}